\documentclass[lettersize,journal]{IEEEtran}
\usepackage{amsmath,amsfonts}
\usepackage{algorithm}
\usepackage{array}
\usepackage[caption=false,font=normalsize,labelfont=sf,textfont=sf]{subfig}
\usepackage{textcomp}
\usepackage{stfloats}
\usepackage{url}
\usepackage{verbatim}
\usepackage{graphicx}
\usepackage{cite}
\usepackage{xspace}
\usepackage{multirow}
\usepackage{booktabs}
\usepackage{algpseudocode} 
\algrenewcommand\algorithmicrequire{\textbf{Input:}}
\algrenewcommand\algorithmicensure{\textbf{Output:}}
\newcommand{\projtitle}{\textsc{HiRL}\xspace}

\begin{document}

\title{\projtitle: Hierarchical Reinforcement Learning for Coordinated Resource Management in Heterogeneous Edge Computing}


\author{
        {Jianyong~Zhu,
        Hao~Chen,
        Juan~Zhang,~\IEEEmembership{Member,~IEEE},
        Fangda Guo,~\IEEEmembership{Member,~IEEE},
        Albert~Y.~Zomaya,~\IEEEmembership{Fellow,~IEEE},
        Renyu~Yang,~\IEEEmembership{Member,~IEEE}}
        \vspace{-0.45cm}

% \thanks{Manuscript received XXX; revised XXX; accepted XXX.This work was supported by the Hebei Provincial Natural Science Foundation (F2023502002) and the Fundamental Research Funds for the Central Universities (2025MS153).\\
% (corresponding author: Renyu Yang)}

\thanks{Received XXX; revised XXX; accepted XXX. Date of publication XXX; date of current version XXX.This work was supported in part by the Hebei Provincial Natural Science Foundation under Grant F2023502002 and in part by the Fundamental Research Funds for the Central Universities under Grant 2025MS153. (Corresponding author: Renyu Yang)}

\IEEEcompsocitemizethanks{
    % \IEEEcompsocthanksitem J. Zhu and H. Chen are with China and North China Electric Power University, Baoding, 071003, China. Email:zhujy@ncepu.edu.cn; chenh@ncepu.edu.cn.
    % \IEEEcompsocthanksitem J.Zhu is with the Yanzhao Electric Power Laboratory of North China Electric Power University, and Hebei Key Laboratory of Knowledge Computing for Energy \& Power. Email: zhujy@ncepu.edu.cn
    % \IEEEcompsocthanksitem H. Chen is with the Department of Computer, North China Electric Power University, and Engineering Research Center of Intelligent Computing for Complex Energy Systems, Ministry of Education. Email: chenh@ncepu.edu.cn.
    % \IEEEcompsocthanksitem  J. Zhang is with the James Watt School of Engineering, University of Glasgow, Glasgow, G12 8QQ, UK. Email: juan.zhang@glasgow.ac.uk.
    % \IEEEcompsocthanksitem R. Yang is with the School of Software, Beihang University, Beijing, 100191, China. Email: renyuyang@buaa.edu.cn.
    \IEEEcompsocthanksitem J. Zhu is with the Yanzhao Electric Power Laboratory of North China Electric Power University, and Hebei Key Laboratory of Knowledge Computing for Energy \& Power, Baoding 071000, China (e-mail: zhujy@ncepu.edu.cn).
    \IEEEcompsocthanksitem H. Chen is with the Department of Computer, North China Electric Power University, and Engineering Research Center of Intelligent Computing for Complex Energy Systems, Ministry of Education, Baoding 071000, China (e-mail: chenh@ncepu.edu.cn).
    \IEEEcompsocthanksitem J. Zhang is with the James Watt School of Engineering, University of Glasgow, Glasgow G12 8QQ, U.K. (e-mail: juan.zhang@glasgow.ac.uk).
   \IEEEcompsocthanksitem F. Guo is with the State Key Laboratory of AI Safety, Institute of Computing Technology, Chinese Academy of Sciences, Beijing 100190, China (Email: guofangda@ict.ac.cn).
    % \IEEEcompsocthanksitem S. Yang is with School of Computer Science and Technology in Beijing Institute of Technology, Beijing, 100191, China (email: S.Yang@bit.edu.cn). 
    \IEEEcompsocthanksitem Albert Zomaya is with the University of Sydney, Australia. E-mail: albert.zomaya@sydney.edu.au.
    \IEEEcompsocthanksitem R. Yang is with the School of Software, Beihang University, Beijing 100191, China (e-mail: renyuyang@buaa.edu.cn).
}

}

%\markboth{Journal of \LaTeX\ Class %Files,~Vol.~14, No.~8, August~2021}%

\markboth {IEEE TRANSACTIONS ON COMPUTERS, VOL. XX, NO. X, JANUARY XXXX}%
% \markboth {\tiny ZHU \MakeLowercase{et al.}: HiRL: HIERARCHICAL REINFORCEMENT LEARNING FOR COORDINATED RESOURCE MANAGEMENT IN HETEROGENEOUS MOBILE EDGE COMPUTING}%
% {\fontsize{6.5pt}{9.2pt}\selectfont ZHU \MakeLowercase{et al.}: \projtitle: HIERARCHICAL REINFORCEMENT LEARNING FOR COORDINATED RESOURCE MANAGEMENT IN HETEROGENEOUS MOBILE EDGE COMPUTING}
{ZHU \MakeLowercase{et al.}: \projtitle: Hierarchical Reinforcement Learning for Coordinated Resource Management in Heterogeneous Edge Computing}


\maketitle

\begin{abstract}
Edge computing faces unprecedented resource orchestration challenges from multi-dimensional heterogeneity across device architectures, diverse task requirements in CPU-intensive, GPU-intensive, I/O-intensive, and dynamic network conditions. The edge environments demand real-time task processing within strict energy budgets, yet conventional approaches struggle with mixed continuous-discrete optimization while meeting deadline and energy constraints.
This paper presents \projtitle, a hierarchical reinforcement learning framework that decomposes complex resource orchestration into coordinated power control and task allocation decisions. Our approach separates continuous power management using the Twin Delayed Deep Deterministic Policy Gradient (TD3) and discrete task placement using Double Deep Q-Network (DDQN), unified through a coordination engine with five-dimensional queue state representation. 
We propose a heterogeneous assessment of resource compatibility with deadline-oriented prioritization and failure-penalized adaptive sampling to enhance decision quality under resource constraints. To improve practical applicability, the framework models comprehensive system dynamics including device mobility, queue congestion patterns, infrastructure heterogeneity, and priority-sensitive scheduling demands.
Experimental results show that \projtitle achieves effective latency-energy trade-offs with 28\% latency reduction compared to Single-DDQN and maintains nearly 100\% task completion rates under all load conditions. Compared to baseline algorithms, \projtitle reduces energy consumption by up to 51\% under low load while achieving 24\% better latency performance than static optimization approaches under high load, establishing effective resource orchestration in heterogeneous edge environments.
\end{abstract}


\section{Introduction}
\label{sec:Introduction}

\IEEEPARstart{E}{dge} computing addresses computational demands of IoT devices with limited resources, dynamic network conditions, and strict energy constraints, particularly in mobile edge computing (MEC) scenarios~\cite{mach2017mobile, nasir2025relto}. The environments face resource orchestration challenges from infrastructure heterogeneity—edge servers and mobile devices differ in CPU/GPU configurations and storage capacities~\cite{zeng2021energy, islam2025delta}—and workload heterogeneity spanning CPU-intensive tasks, GPU-accelerated applications, and I/O-intensive transfers~\cite{lee2022real, kim2020gpu}. Device mobility further complicates scheduling through time-varying network conditions~\cite{akhlaqi2023task, wang2024joint}. Ineffective resource management leads to deadline violations, energy waste, and task failures.

\begin{figure}[t]
\centering
\includegraphics[width=0.48\textwidth]{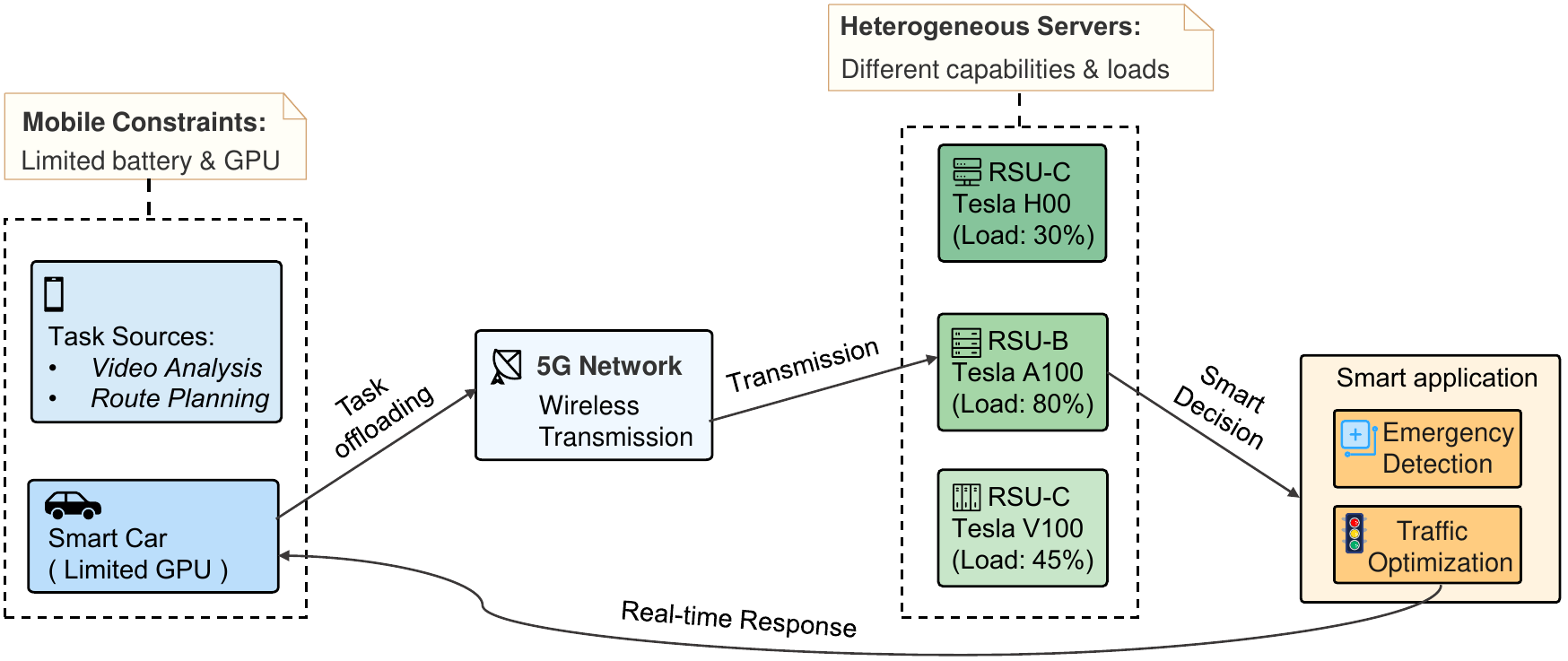}
\caption{Intelligent transportation scenario: Smart vehicle offloads video analysis and route planning tasks to heterogeneous edge servers.}
\label{fig:scenario}
\vspace{-1em}
\end{figure}

Current approaches face three limitations. First, independent handling of power control and task allocation prevents joint optimization—existing methods either fix transmission power during task allocation~\cite{hou2022optimal, pan2021multi} or separate frequency scaling from offloading decisions~\cite{kalpana2024deep, islam2025delta}, missing energy-latency tradeoffs. Second, treating GPU-intensive and CPU-intensive tasks uniformly causes resource mismatches: GPU tasks on CPU-only servers fail, while CPU tasks on GPU servers waste resources~\cite{cao2024dependent, chen2023dynamic}. Third, optimizing energy, latency, or completion rates separately creates conflicts—aggressive offloading reduces local energy but increases transmission and congestion costs, worsening latency and deadline misses~\cite{zeng2025task, dai2023uav}. 

Consider a computing scenario in Fig.~\ref{fig:scenario}. A vehicle offloads GPU-intensive video analysis and CPU-intensive route planning to heterogeneous servers (H100/A100/V100) under adjustable transmission power (up to 3W) with emergency deadlines. Independent decisions fail: high power without GPU availability wastes energy; GPU tasks on incompatible servers violate deadlines. Effective management requires hierarchical coordination with GPU-aware matching.

We present \projtitle, a hierarchical reinforcement learning framework for resource orchestration in heterogeneous MEC environments. The framework addresses mixed continuous-discrete optimization through hierarchical decomposition while maintaining cross-tier coordination. We design a dual-pipeline CPU/GPU computational model to represent heterogeneous resource requirements. The two-tier architecture separates continuous power control (upper tier TD3 with five-dimensional queue state managing CPU frequency and transmission power) from discrete task allocation (lower tier DDQN with GPU-aware state design for placement across heterogeneous servers), unified through a coordination engine implementing a three-stage decision pipeline. We introduce deadline-oriented queue management and failure-penalized experience replay to handle real-time constraints and accelerate learning from resource violations.

We evaluate \projtitle across production-scale heterogeneous deployments with 35 mobile devices and 5 edge servers under controlled load variations. Experimental results show 28\% latency reduction compared to Single-DDQN while maintaining 98\%+ task completion rates. Energy consumption decreases by 51\% under low load, and latency improves by 24\% over static optimization under high load. Ablation studies confirm hierarchical coordination and GPU-aware compatibility assessment contribute 49-82\% of these gains, while disabling failure-penalized learning causes system collapse with completion rates dropping to 62\%.

The main contributions include:
\begin{itemize}
    \item Hierarchical framework with coordinated CPU/GPU dual-pipeline modeling for heterogeneous MEC environments.
    \item Compatibility-aware task allocation mechanism with GPU-resource matching across heterogeneous servers.
    \item Deadline-oriented queue optimization and failure-penalized sampling for constraint handling.
\end{itemize}

The remainder is organized as follows: Section~\ref{sec:hierarchical} presents framework and modeling; Sections~\ref{sec:power_control} and~\ref{sec:task_allocation} detail TD3 power control and DDQN task allocation; Section~\ref{sec:evaluation} provides experimental evaluation; Section~\ref{sec:related} reviews related work; Section~\ref{sec:conclusion} concludes. Source codes and data are available at https://github.com/ch-ncepu/HIRL.

\section{Hierarchical Framework and System Modeling}
\label{sec:hierarchical}

\subsection{Motivation}
\label{subsec:motivation}
Heterogeneous MEC environments present resource orchestration challenges coupling continuous power control and discrete task placement under stringent deadlines. In the vehicular scenario of Fig.~\ref{fig:scenario}, emergency video analysis requires 4~GB GPU memory and 2048 CUDA cores within 200~ms, whereas route planning demands 50~M CPU cycles with negligible GPU use. The vehicle adjusts transmission power (0.1–3~W) and CPU frequency (0.8–2.4~GHz) while selecting among heterogeneous servers (H100/A100/V100) under LTE-V2X constraints.

This scenario exposes key optimization challenges. The joint decision space combines continuous control with discrete selection: for $M$ devices choosing among $N$ servers, the space grows as $O((N+1)^M \times \mathbb{R}^{2M})$. Discretizing power into 10 frequency and 10 power levels yields 100 combinations per device, reducing precision. Relaxing discrete choices to continuous variables produces non-convex landscapes where gradient-based optimization becomes unstable.

Power decisions and task placement are dynamically interdependent. Increasing transmission power from 0.5W to 2W shortens offloading latency $D_{tx}^i$, enabling GPU execution for latency-critical tasks. Conversely, frequency scaling alters local processing: a task missing its deadline at 0.8GHz may complete locally at 2.4GHz with lower energy cost. Decoupled optimization breaks these dependencies—power control without task awareness may select conservative transmission, causing deadline violations, while allocation without power constraints may offload to distant servers, causing excessive latency.

% Task heterogeneity further amplifies these dependencies by introducing coupled queue dynamics that single-queue models cannot represent. GPU-intensive video analysis, CPU-bound route planning, and I/O-driven data transfer generate distinct workloads across heterogeneous resources. We capture these interactions using five queue dimensions: local CPU, local GPU, wireless transmission, server CPU, and server GPU. Each queue accumulates backlog differently according to task mix and processing capacity—CPU queues grow steadily with general arrivals, GPU queues surge with bursty vision tasks, and transmission queues fluctuate with channel conditions and data size. Modeling these interdependent queues jointly with power-allocation decisions requires learning over a high-dimensional state space, leading to sample inefficiency and unstable convergence in dynamic environments.
Task heterogeneity introduces coupled queue dynamics across heterogeneous resources. GPU-intensive video analysis, CPU-bound route planning, and I/O-driven transfers create distinct workload patterns. We model these interactions through five queue dimensions: local CPU, local GPU, wireless transmission, server CPU, and server GPU. Queue backlogs evolve differently—CPU queues grow steadily with general arrivals, GPU queues surge with bursty vision tasks, transmission queues fluctuate with channel conditions. Joint optimization of these interdependent queues with power-allocation decisions requires learning over high-dimensional state spaces, causing sample inefficiency in dynamic environments.

\begin{figure}[t]
\centering
\includegraphics[width=0.48\textwidth]{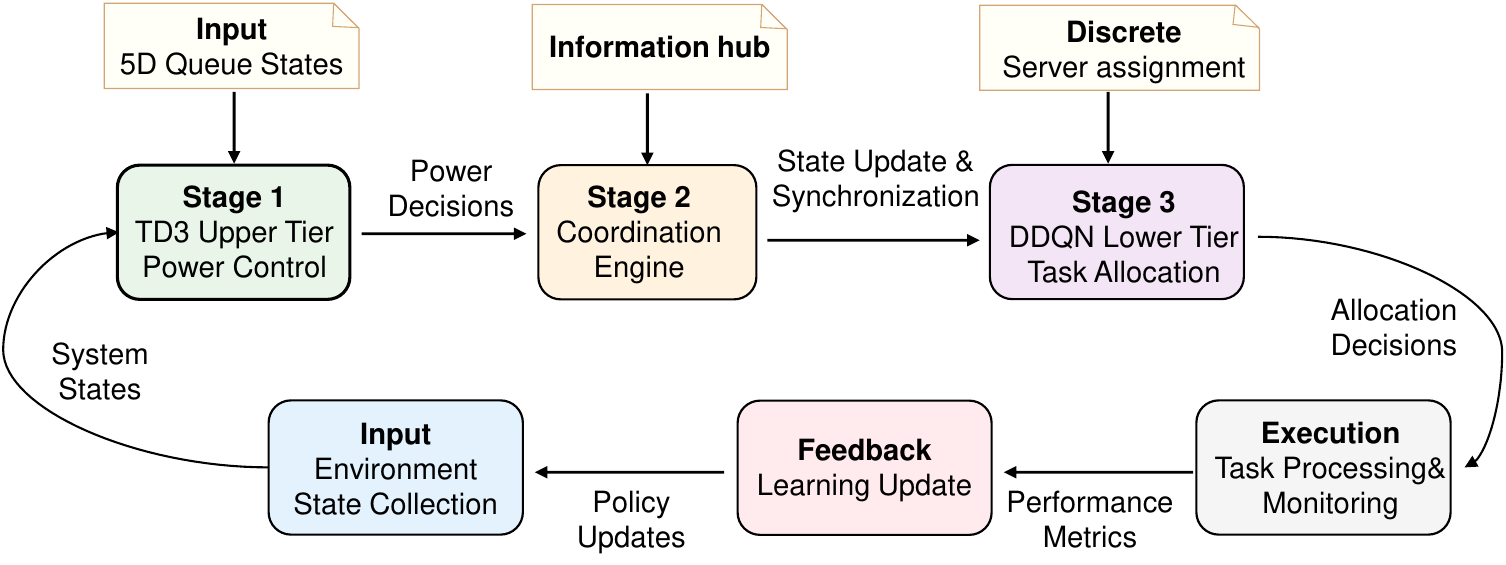}
\caption{Workflow of \projtitle: Synchronized three-stage 
decision pipeline with closed-loop feedback.}
\label{fig:workflow}
\vspace{-0.35cm}
\end{figure}

We address these challenges through a hierarchically coordinated framework (Fig.~\ref{fig:workflow}). The upper tier employs TD3 for continuous power control, adjusting $(f_l^t, p^t)$ based on queue states $\mathbf{q}^t$ and channel gains $g^t$. The lower tier uses DDQN for discrete tasks offloading, integrating GPU compatibility and transmission rate updates. A three-stage pipeline links the tiers: Stage~1 determines power settings, Stage~2 updates transmission rates and queue states, Stage~3 finalizes task allocation. This design decouples optimization objectives while preserving inter-tier dependencies through shared queues.

\subsection{Problem Formulation}

Consider the vehicle scenario from Fig.~\ref{fig:scenario}. GPU-intensive video analysis requires substantial CUDA cores and memory, while CPU-intensive route planning demands computational cycles. Both scenarios face stringent emergency deadlines and constrained transmission power (2-3W), and thus come with a resource orchestration problem -- coordinating power management with task placement across $M$ mobile devices and $N$ edge servers ($M=35$, $N=5$ in our evaluation) to process three task categories $\tau = \{t_{\text{CPU}}, t_{\text{GPU}}, t_{\text{IO}}\}$.

Each task $i$ has attributes $(b_i, s_i, c_i, ddl_i, n_i, m_i, s\_gpu_i,\\ t_{create})$ representing the priority, data size, CPU cycles, deadline, required CUDA cores, GPU memory, and computational load and the creation time of the task. The execution cost (Eq.~\ref{eq:cost}) strike a balance between latency and energy. 
\begin{equation}
cost_i = \lambda b_i D_{\text{tot}}^i + (1 - \lambda) E^i,
\label{eq:cost}
\end{equation}
where $b_i$ amplifies the weight for critical tasks and $\lambda \in [0,1]$ controls the tradeoff. The total latency $D_{\text{tot}}^i$ aggregates queuing latencies across local/transmission/server stages and CPU-GPU pipeline execution. The energy $E^i$ combines local processing $\kappa (f_l^t)^2 c_i$, transmission $p^t D_{tx}^i$, and server-side GPU consumption.

The optimization objective (Eq.~\ref{eq:min_cost}) minimizes long-term execution costs through coordinated power control and task allocation decisions. 
\begin{equation}
\begin{aligned}
\min \quad & \lim_{T \to \infty} \frac{1}{T}\sum_{t=0}^{T}\sum_{i \in \Gamma_t} cost_i, \\
\text{s.t.} \quad & 0 \leq f_t^j \leq f_{max}^j, \; 0 \leq p_t^j \leq p_{max}^j, \quad \forall j, t, \\
& D_{\text{tot}}^i \leq ddl_i, \quad \forall i, \\
& \sum_{i \in \mathcal{A}_j} n_i \leq N_{gpu}^j, \; \sum_{i \in \mathcal{A}_j} m_i \leq M_{gpu}^j, \quad \forall j, 
\end{aligned}
\label{eq:min_cost}
\end{equation}
where $\Gamma_t$ denotes the set of tasks arriving at time $t$, $\mathcal{A}_j$ represents the tasks assigned to server $j$, and GPU-related constraints are enforced at the server level through the available CUDA cores $N_{gpu}^j$ and memory capacity $M_{gpu}^j$. The resulting resource orchestration problem involves tightly coupled \textit{continuous decisions}, including CPU frequency allocation $f_t^j$ and transmission power control $p_t^j$, together with \textit{discrete decisions} on task placement between local execution and edge servers. The proposed hierarchical framework addresses this mixed-integer optimization by decomposing it into coordinated subproblems, which are jointly optimized in accordance with Eq.~\ref{eq:min_cost}. Table~\ref{tab:notation} summarizes all notations used throughout the framework.
% \noindent where $\Gamma_t$ denotes tasks arriving at time $t$, $\mathcal{A}_j$ represents tasks allocated to server $j$, and GPU constraints apply per server with CUDA cores $N_{gpu}^j$ and memory $M_{gpu}^j$. The resource orchestration problem involves continuous decisions (CPU frequency $f_t^j$ and transmission power $p_t^j$) coupled with discrete decisions (task placement across local execution and edge servers). Our hierarchical approach decomposes this hybrid optimization into coordinated sub-problems toward the global objective.

% The resource orchestration problem spans both continuous and discrete decision spaces:
% \begin{itemize}
% \item \textit{Continuous domain:} CPU frequency adjustment $f_t^j$ and transmission power control $p_t^j$.
% \item \textit{Discrete domain:} Task placement (local vs. edge execution) and server resource selection.
% \end{itemize}

% Our hierarchical framework decomposes this mixed-domain optimization into coordinated subproblems aligned with the global objective in Eq.~\ref{eq:min_cost}.

\begin{table}[t]
\centering
\caption{Notation}
\label{tab:notation}
\scriptsize
\begin{tabular}{c|p{6.2cm}}
% \toprule
\hline
\textbf{Symbol} & \textbf{Description} \\
% \midrule
\hline
$M$, $N$ & Mobile devices, edge servers \\
$\mathbf{q}^t$ & Five-dimensional queue state \\
$f_l^t$, $p^t$ & CPU frequency, transmission power \\
$g^t$, $r^t$ & Channel gain, transmission rate \\
$\rho_i^j$ & GPU compatibility score \\
$\mathcal{T}_i$ & Task attributes 
% $(b_i, s_i, c_i, ddl_i, n_i, m_i, g_i)$ 
$(b_i, s_i, c_i, ddl_i, n_i, m_i, s\_gpu_i, t_{create})$\\
$\lambda$ & Energy-latency tradeoff weight \\
$D_{tot}^i$, $E^i$ & Task latency, energy \\
% \bottomrule
\hline
\end{tabular}
\end{table}

\subsection{Hierarchical Decision Framework}
\label{subsec:hierarchy}

The hierarchical architecture decomposes the hybrid optimization problem (Eq.~\ref{eq:min_cost}) into coordinated submodules. The upper tier applies TD3 for continuous power control ($f_t^j, p_t^j$), mitigating value overestimation through twin critics and delayed updates. The lower tier adopts DDQN for discrete task placement, decoupling action selection from value evaluation for GPU-heterogeneous server choices. Both tiers interact through a coordination engine synchronizing decisions via a three-stage pipeline (Fig.~\ref{fig:workflow}).

At each time slot $t$ ($\tau=1$s), the system executes synchronized operations across all $M$ devices. The upper tier first updates the five-dimensional queue state $\mathbf{q}^t$, capturing local CPU/GPU, transmission, and server CPU/GPU backlogs, while measuring channel gains $g^t$ to generate frequency–power pairs $(f_t^j, p_t^j)$. The coordination engine then updates transmission rates $r^t$ based on the new power settings and channel conditions, normalizes queue states relative to current capacities, and forwards the updated states to the lower tier. The lower tier determines server placements for new tasks according to GPU compatibility and deadline urgency. Task dispatch follows deadline-oriented sorting, with execution outcomes generating reward signals $r_p^t$ and $r_o^i$ for policy refinement.

The coordination engine maintains consistency across tiers through three mechanisms via Eq.~\ref{eq:pipeline}:

\textbf{Shared State Representation}. Both tiers operate on a synchronized five-dimensional queue vector $\mathbf{q}^t$ (Eq.~\ref{eq:5d}), providing a common view of resource utilization across heterogeneous components.

\textbf{Sequential Decision Pipeline}. The three-stage structure creates explicit tier dependencies:
\begin{equation}
\begin{aligned}
\text{Stage 1: } &(f^t, p^t) = \pi_p(\mathbf{q}^t, g^t), \\
\text{Stage 2: } &r^t = \text{Update}(p^t, g^t), \quad \hat{\mathbf{q}}^t = \text{Normalize}(\mathbf{q}^t, f^t, r^t), \\
\text{Stage 3: } &a_o^i = \pi_o(\hat{\mathbf{q}}^t, f^t, p^t, \mathcal{T}_i, r^t),
\end{aligned}
\label{eq:pipeline}
\end{equation}
\noindent where TD3 policy $\pi_p$ tunes frequency-power pairs; Stage~2 updates transmission rates and normalized queues; DDQN policy $\pi_o$ selects task placements using refreshed states. The pipeline maintains causal consistency—power control directly determines feasible allocation boundaries.

\textbf{Bidirectional Feedback.} Execution outcomes are fed back via coordinated rewards to jointly refine both policies and align them with global objectives (Sections~\ref{sec:power_control} and~\ref{sec:task_allocation}).

\subsection{State Representation}
\label{subsec:state}
The framework uses a five-dimensional queue state (Eq.~\ref{eq:5d}) tracking resource utilization across system components:
\begin{equation} 
\mathbf{q}^t = [q_{lc}^t, q_{tx}^t, q_{lg}^t, q_{sc}^t, q_{sg}^t] ,
\label{eq:5d} 
\end{equation}
\noindent where subscripts $lc$, $sc$ denote local/server CPU queues, $lg$, $sg$ represent local/server GPU queues, and $tx$ indicates transmission queue. Each queue accumulates workload from waiting tasks, where $\mathcal{Q}_{j}^t$ represents tasks in queue $j$ at time slot $t$: $q_{lc}^t = \sum_{i \in \mathcal{Q}_{lc}^t} w_i$, $q_{tx}^t = \sum_{i \in \mathcal{Q}_{tx}^t} s_i$, $q_{sc}^t = \sum_{i \in \mathcal{Q}_{sc}^t} w_i$, $q_{lg}^t = \sum_{i \in \mathcal{Q}_{lg}^t} s\_gpu_i$, $q_{sg}^t = \sum_{i \in \mathcal{Q}_{sg}^t} s\_gpu_i$, with $w_i = s_{cpu,i} \times c_i$ denoting required CPU cycles and $s\_gpu_i$ denoting GPU computational load (FLOPS). Queue evolution couples upper-tier power control with lower-tier task allocation through task arrivals and processing departures.

The system load metric (Eq.~\ref{eq:load_metric}) integrates queue backlogs with real-time utilization:
\begin{equation}
L^t = \frac{\omega_q}{5} \sum_{j \in \mathcal{Q}} \tanh(\hat{q}_j^t) + \omega_u \cdot u_{cpu}^t + \omega_g \cdot u_{gpu}^t,
\label{eq:load_metric}
\end{equation}
\noindent where $\mathcal{Q} = \{lc, tx, lg, sc, sg\}$ and weights $\omega_q$, $\omega_u$, $\omega_g$ balance queue backlog with CPU/GPU utilization.

Queue normalization scales states relative to processing capacity through Eq.~\ref{eq:normalized}:
\begin{equation}
\hat{q}_{j}^t = \frac{q_{j}^t}{f_j^t \cdot \tau}, \quad j \in \{lc, tx, lg, sc, sg\}
\label{eq:normalized},
\end{equation}
\noindent where $f_j^t$ denotes processing rate for queue $j$ and $\tau$ is time slot duration. Values $\hat{q}_{j}^t > 1$ indicate backlog exceeding single-slot capacity.

\textbf{Tier-Specific States.} The upper tier observes state vector $S_p$ defined in Eq.~\ref{eq:upper}:
\begin{equation}
S_p = [\mathbf{q}^t, g^t, f_{max}, p_{max}],
\label{eq:upper}
\end{equation}
\noindent where $g^t$ denotes channel gain and $f_{max}$, $p_{max}$ denote capability bounds. The lower tier uses extended state $S_o$ (Eq.~\ref{eq:lower}):
\begin{equation}
\begin{aligned}
S_o = \big[
&q_{lc}^t, q_{tx}^t, q_{lg}^t, q_{sc}^t, q_{sg}^t, f^t, p^t, g^t, \\
&gpu_f, gpu_{num}, gpu_{bw}, gpu_{pwr}, \\
&u_{cpu}^{l,t}, u_{gpu}^{l,t}, u_{cpu}^{s,t}, u_{gpu}^{s,t}, \\
&\mathcal{T}_i(b_i, s_i, c_i, ddl_i, n_i, m_i, s\_gpu_i, t_{create})
\big],
\end{aligned}
\label{eq:lower}
\end{equation}
\noindent where GPU specifications $(gpu_f, gpu_{num}, gpu_{bw}, gpu_{pwr})$ represent frequency, CUDA cores, bandwidth, and power consumption; utilization metrics $(u_{cpu}^{l,t}, u_{gpu}^{l,t}, u_{cpu}^{s,t}, u_{gpu}^{s,t})$ track local/server CPU/GPU usage; task tuple $\mathcal{T}(\cdot)$ contains priority, size, complexity, deadline, GPU requirements, and creation timestamp. DeDetails are in Sections~\ref{sec:power_control} and~\ref{sec:task_allocation}.

% \subsection{Coordinated Algorithm}
% \label{subsec:coordination_alg}

Alg.~\ref{alg:hermes_coordination} integrates the hierarchical framework through synchronized three-stage operations per time slot.

\begin{algorithm}[t]
\caption{\projtitle~Hierarchical Coordination.}
\footnotesize
\label{alg:hermes_coordination}
\begin{algorithmic}[1] % <-- [1] 启用自动行号（algpseudocode 特性）
\Require Devices $\mathcal{M}$, servers $\mathcal{N}$, TD3 agent $\pi_p$, DDQN agent $\pi_o$
\State $t \gets 1$
\While{$t \leq T$}
    \State {\textbf{// Stage 1: Power Control}}
    \For{$j \in \mathcal{M}$}
        \State Collect queue state: $\mathbf{q}_j^t$, measure channel gain: $g_j^t$
        \State Execute power control: $(f_l^t, p^t) \gets \pi_p(\mathbf{q}_j^t, g_j^t)$
    \EndFor

    \State {\textbf{// Stage 2: State Synchronization}}
    \State Update transmission rates: $r^t \gets \text{ComputeRate}(p^t, g^t)$
    \State Normalize queue states using Eq.~\ref{eq:normalized}
    \State Update load metric $L^t$ using Eq.~\ref{eq:load_metric}

    \State {\textbf{// Stage 3: Task Allocation}}
    \For{each arriving task $i$ on device $j$}
        \State Extract task attributes: $\mathcal{T}_i$
        \State Compute compatibility: $\rho_i^k$ for $k \in \mathcal{N}\cup\{0\}$
        \State Execute allocation: $a_o^i \gets \pi_o(\mathbf{q}_j^t, f_l^t, p^t, \mathcal{T}_i, r^t)$
        \State Dispatch task to selected queue
    \EndFor

    \State {\textbf{// Execution and Learning}}
    \State Process tasks with deadline prioritization
    \State Collect outcomes: completion status, latencies, energy
    \State Calculate rewards: $r_p^t$ (Eq.~\ref{eq:power_reward}), $r_o^i$ (Eq.~\ref{eq:task_reward})
    \State Store experiences with failure-aware prioritization
    \State Update TD3 and DDQN networks

    \State $t \gets t + 1$
\EndWhile
\end{algorithmic}
\end{algorithm}

% Time-slot synchronization ensures consistent state representation across heterogeneous devices. Updated capacities link power control to task allocation, and execution feedback jointly refines both tiers toward global optimality. Detailed reward formulations and training mechanisms are presented in Sections~\ref{sec:power_control} and~\ref{sec:task_allocation}.
Time-slot synchronization maintains consistent states across tiers. Detailed reward formulations are in Sections~\ref{sec:power_control} and~\ref{sec:task_allocation}.

\section{Power Control with Twin Delayed Deep Deterministic Policy Gradient}
\label{sec:power_control}

This section presents the upper-tier TD3-based power control mechanism within the hierarchical framework in Section~\ref{sec:hierarchical}, including problem formulation, state-action representation, reward design, and implementation details.

\subsection{TD3 for Power Control in Heterogeneous Environments}
\label{subsec:td3_power_control}

The upper tier employs TD3 to optimize continuous power control under heterogeneous MEC settings. TD3 incorporates twin critics to mitigate value overestimation, delayed policy updates to suppress oscillations, and target smoothing to improve training stability across diverse device capabilities and fluctuating network conditions.

% \textbf{Power Control Formulation.} TD3 operates on state $S_p^t$ (Eq.~\ref{eq:upper}) and action space $A_p = [f_l^t, p^t]$ with device-specific constraints, optimizing:
% \begin{equation}
% \max_{\pi_p} \mathbb{E}\left[\lim_{T \to \infty} \frac{1}{T}\sum_{t=0}^{T} r_p^t\right]
% \label{eq:td3_objective}
% \end{equation}
% \noindent where $\pi_p$ represents the power control policy and $r_p^t$ follows the coordinated reward design from Eq.~\ref{eq:power_reward}. TD3 learns policies aligned with global system performance while respecting device-specific constraints.
\textbf{Power Control Formulation.} TD3 operates on state $S_p^t$ (Eq.~\ref{eq:upper}) and action space $A_p = [f_l^t, p^t]$ with device-specific constraints, optimizing $\max_{\pi_p} \mathbb{E}[\lim_{T \to \infty} \frac{1}{T}\sum_{t=0}^{T} r_p^t]$, where $\pi_p$ denotes the power control policy and $r_p^t$ is defined in Eq.~\ref{eq:power_reward}.

\subsection{State-Action Representation and Normalization}
\label{subsec:power_state_action}

\textbf{State Space Components.} The power control agent processes state space $S_p^t$ (Eq.~\ref{eq:upper}). Channel gain follows the path-loss model in Eq.~\ref{eq:channel_gain}:
\begin{equation}
g^t = G_0 \left(\frac{d_0}{d^t}\right)^\gamma,
\label{eq:channel_gain}
\end{equation}
\noindent where $G_0$ is the gain constant, $d^t$ denotes the distance, and $\gamma$ is the path-loss exponent affecting transmission rates in Stage~2 (Eq.~\ref{eq:pipeline}).

\textbf{Synchronized State Management:} State updates are synchronized at time-slot boundaries, such that power-control outputs from Stage~1 are immediately applied to Stage~3 allocation, ensuring TD3 and DDQN operate on a consistent system state.

\subsubsection{Continuous Action Space Formulation}

The power control agent outputs continuous actions within a constrained space as defined in Eq.~\ref{eq:action_space_power}:
\begin{equation}
A_p = [f_l^t, p^t], \quad f_l^t \in [0, f_{max}], \quad p^t \in [0, p_{max}],
\label{eq:action_space_power}
\end{equation}

\noindent where $f_l^t$ denotes the local CPU frequency (Hz) and $p^t$ the transmission power (W). 
These actions implement Stage~1 (Eq.~\ref{eq:pipeline}): $f_l^t$ determines local processing capacity reflected in queue states, while $p^t$ determines the communication rate $r^t$ in Stage~2, both of which are included in the task allocation agent’s state.

\textbf{State Normalization.} Queue states are normalized following Eq.~\ref{eq:normalized}: $\hat{q}_{lc}^t = q_{lc}^t/(f_l^t \cdot \tau)$ to ensure scale-invariant learning across varying device configurations and workload intensities.

\subsection{Reward Function and Training Optimization}
\label{subsec:power_reward_training}

The hierarchical framework aligns both tiers toward global objectives through coordinated reward structures. The upper tier optimizes system-level performance (Eq.~\ref{eq:power_reward}):
\begin{equation}
r_p^t = R_{complete}^t - \lambda_e \cdot E^t, 
\label{eq:power_reward}
\end{equation}
\noindent where $R_{complete}^t = \sum_{i \in \mathcal{C}^t} b_i I_{success}^i$ sums priority-weighted completed tasks ($I_{success}^i \in \{0,1\}$), and $E^t$ aggregates CPU and transmission energy weighted by $\lambda_e$, assigning higher priority to task completion than energy reduction.

The lower tier optimizes individual task execution via Eq.~\ref{eq:task_reward}:
\begin{equation}
r_o^i = -(\lambda b_i D_{tot}^i + (1-\lambda) E^i) - \alpha \cdot I_{fail}^i,
\label{eq:task_reward}
\end{equation}
\noindent where the first term matches Eq.~\ref{eq:cost}, ensuring tier alignment. Penalty $\alpha$ discourages constraint violations indicated by $I_{fail}^i \in \{0,1\}$ through GPU compatibility checks. Both rewards share metrics ($D_{tot}^i$, $E^i$, completion status), maintaining feedback consistency along the pipeline (Eq.~\ref{eq:pipeline}).

% The TD3 agent performance in our heterogeneous environment depends on a reward structure that captures the energy-performance trade-off and training methods for multi-dimensional queue dynamics.

% \subsubsection{Energy-Queue Balancing Reward}

% The power control agent uses the coordinated reward design from Eq.~\ref{eq:power_reward}, where task completion success directly drives policy optimization. The energy penalty term ensures efficient resource usage without compromising system functionality.

% \begin{equation}
% E^t = k(f_l^t)^2\tau + p^t \frac{s_{tx}^t}{r^t}
% \label{eq:energy_components}
% \end{equation}

% \noindent where $k$ is the effective switching capacitance, $f_l^t$ is the local CPU frequency from Eq.~\ref{eq:action_space_power}, $s_{tx}^t$ is the transmitted data size, and $r^t$ is the transmission rate updated using the channel gain from Eq.~\ref{eq:channel_gain}. The energy calculation in Eq.~\ref{eq:energy_components} feeds directly into the coordinated reward structure, ensuring that power control decisions support global system objectives through the adaptive weight mechanism. This formulation naturally balances immediate energy savings against queue congestion across heterogeneous resources.

%\subsubsection{Training Optimization and Energy-Completion Balance}
\subsubsection{Energy-Performance Trade-off Optimization}

The TD3 agent implements the coordinated reward from Eq.~\ref{eq:power_reward} with energy computation following Eq.~\ref{eq:energy_components}:
\begin{equation}
E^t = k(f_l^t)^2\tau + p^t \frac{s_{tx}^t}{r^t}, 
\label{eq:energy_components}
\end{equation}
\noindent where $k$ denotes the effective switching capacitance, $(f_l^t)^2$ reflects the quadratic CPU frequency--power relationship, and transmission energy depends on data size $s_{tx}^t$ and communication duration. 

The fixed coefficient $\lambda_e = 0.5$ in Eq.~\ref{eq:power_reward} assigns higher weight to task completion, yielding over 98\% completion rates; during training, TD3 accordingly allocates power aggressively for high-priority, deadline-critical tasks and conservatively under low load, driven solely by the reward structure, without explicit state-dependent weighting.

\subsubsection{Failure-Aware Experience Prioritization}

Our training method prioritizes experiences from critical scenarios to accelerate learning in challenging situations. The prioritization weight combines temporal difference error with failure indicators as shown in Eq.~\ref{eq:failure_priority}:
\begin{equation}
W_i = |TD_i| \cdot (1 + \gamma_{fail} \cdot I_{fail}^i), 
\label{eq:failure_priority}
\end{equation}
\noindent where $|TD_i|$ denotes the temporal difference error magnitude, $I_{fail}^i$ indicates task failures, and $\gamma_{fail}$ amplifies critical scenarios. Prioritization in Eq.~\ref{eq:failure_priority} guides TD3 to focus on system-stress cases induced by power control actions in Eq.~\ref{eq:action_space_power}.

% \textcolor{red}{The reward structure from Eq.~\ref{eq:power_reward} directly incentivizes task completion while penalizing excessive energy consumption. Unlike adaptive weighting schemes that may inadvertently sacrifice task completion for energy savings, the fixed penalty coefficient $\lambda_e$ ensures consistent prioritization of system functionality.}

\subsection{Network Architecture and Implementation}
\label{subsec:td3_architecture}

The TD3 implementation for power control uses network architectures and training strategies for dynamic heterogeneous mobile edge environments.

\subsubsection{Actor-Critic Network Design}
%----------TODO----------
%TD3两个c网络不准确，应该是2个Critic,2个目标cirtic
The TD3 framework consists of an actor network and twin critic networks with target stabilization for continuous power control. The actor $\mu_\phi(s)$ maps the six-dimensional state $S_p^t$ to the two-dimensional action $a^t = [f_l^t, p_l^t]$ using a tanh output layer, followed by reverse normalization based on state-provided bounds $(f_{max}, p_{max})$ to scale actions from $[-1,1]$ to device-specific ranges. Through this scaling mechanism, a shared TD3 policy is applied across heterogeneous devices.

Complementing the actor, the twin critics $Q_{\theta_1}(s,a)$ and $Q_{\theta_2}(s,a)$ adopt conservative value estimation, $Q_{target} = \min(Q_{\theta_1'}(s', \tilde{a}'), Q_{\theta_2'}(s', \tilde{a}'))$, where $\tilde{a}'$ denotes the smoothed target action, stabilizing value evaluation under mobile edge uncertainty.

\subsubsection{Adaptive Training Mechanisms}

The TD3 implementation incorporates two adaptations for heterogeneous mobile edge environments. 
First, load-adaptive exploration noise is applied as $\sigma^t = \sigma_{base} (1 + \alpha L^t)$, where the load metric $L^t \in [0,1]$ is defined in Eq.~\ref{eq:load_metric} and $\alpha$ controls adaptation sensitivity. This mechanism promotes higher exploration under low-load conditions while encouraging conservative behavior during high-load periods.

Second, the experience replay module applies failure-aware prioritization defined in Eq.~\ref{eq:failure_priority}, emphasizing critical or constraint-violating samples. Together with adaptive exploration noise, this design supports a balanced trade-off between stability and responsiveness under heterogeneous MEC conditions.

\subsubsection{Network Update Mechanism}

% The training process follows a coordinated update schedule adapted for our synchronous time-slot framework. Critic networks are updated at each time slot using standard temporal difference learning with target smoothing, while the actor network employs delayed updates every $d$ time slots to ensure policy stability:
% \begin{equation}
% \label{eq:actor_update}
% \theta_\mu \leftarrow \theta_\mu + \alpha_\mu \nabla_{\theta_\mu} Q_{\theta_1}(s, \mu_{\theta_\mu}(s))
% \end{equation}

Training follows a coordinated update schedule aligned with the synchronous time-slot framework. 
The critic networks are updated at each time slot using temporal difference learning with target smoothing, while the actor network applies delayed updates$\theta_\mu \leftarrow \theta_\mu + \alpha_\mu \nabla_{\theta_\mu} Q_{\theta_1}(s, \mu_{\theta_\mu}(s))$ every $d$ time slots to enhance policy stability.

Target networks are updated via Polyak averaging with parameter $\tau$, ensuring gradual policy evolution across heterogeneous device configurations and dynamic workload conditions.

% Through synchronized updates and soft target evolution, the TD3 agent maintains stable power control policies that adapt gradually to heterogeneous device configurations and dynamic workload patterns.

\section{Task Allocation with Double Deep Q-Network}
\label{sec:task_allocation}
This section presents the lower-tier DDQN-based task allocation mechanism coordinating with the upper-tier power control. It formulates the discrete task allocation problem, describes GPU-aware state–action design, and introduces deadline-oriented queue management with adaptive experience replay. The interaction between DDQN and TD3 within the three-stage execution pipeline is also explained.

\subsection{Double Deep Q-Network for Task Allocation}
\label{subsec:ddqn_task_allocation}

The lower tier employs DDQN for discrete task placement across heterogeneous resources. By decoupling action selection from evaluation, DDQN mitigates value overestimation through target value $Q_{target} = r + \gamma Q_{\theta'}(s', \arg\max_{a'} Q_\theta(s', a'))$, where $Q_\theta$ selects optimal actions and $Q_{\theta'}$ evaluates values, avoiding overestimation when a single network handles both roles. This separation proves advantageous in heterogeneous MEC environments where server diversity and fluctuating loads challenge estimation stability.

\subsubsection{DDQN Implementation in Hierarchical Framework}

DDQN operates as Stage 3 in the coordination pipeline (Eq.~\ref{eq:pipeline}), processing power-updated states with failure penalty rewards (Eq.~\ref{eq:task_reward}). Load-aware exploration follows Eq.~\ref{eq:epsilon_adaptive}:
\begin{equation}
\epsilon^t = \epsilon_{min} + (\epsilon_{max} - \epsilon_{min}) \cdot \exp(-\beta \cdot L^t),
\label{eq:epsilon_adaptive}
\end{equation}
\noindent where $\beta > 0$ controls adaptation sensitivity based on load metric $L^t$ (Eq.~\ref{eq:load_metric}). During high-load periods ($L^t \to 1$), exploration decreases to favor proven strategies, ensuring stability. Low-load conditions enable increased exploration to discover improved policies. This balances exploitation-exploration based on real-time system stress.

\subsection{State Space Design and GPU-Aware Mechanisms}
\label{subsec:ddqn_state_action}

\subsubsection{Multi-dimensional State Representation}

The DDQN agent utilizes the state space from Eq.~\ref{eq:lower} established in Section~\ref{sec:hierarchical}, incorporating the five-dimensional queue state from Eq.~\ref{eq:5d} with power control outputs and task attributes.
\begin{itemize}
\item Five-dimensional queue states $\mathbf{q}^t$ (Eq.~\ref{eq:5d})
\item Power control decisions $(f_l^t, p^t)$ from Stage~1 (Eq.~\ref{eq:pipeline})
\item Task attributes $\mathcal{T}_i 
% (b_i, s_i, c_i, ddl_i, n_i, m_i, g_i)$
(b_i, s_i, c_i, ddl_i, n_i, m_i, s\_gpu_i, t_{create})$
including GPU requirements
\item Network conditions $g^t$ and GPU specifications (frequency, cores, bandwidth, power)
\item Real-time CPU/GPU utilization metrics (local and server)
\item Creation timestamps for latency tracking
\end{itemize}

To manage state dimensionality, we apply normalization from Eq.~\ref{eq:normalized}, providing scale-invariant indicators of resource saturation for consistent policy learning.

\subsubsection{GPU-Aware Resource Matching}

Resource compatibility assessment employs GPU-aware scoring mechanisms to ensure feasible task-resource matching. The compatibility score is defined as follows:
\begin{equation}
\rho_i^j = \min\left(\frac{s_{cpu,i}}{M_{cpu}^j}, \frac{n_i}{N_{gpu}^j}, \frac{m_i}{M_{gpu}^j}\right),
\label{eq:gpu_compatibility}
\end{equation}
\noindent where $\rho_i^j \in [0,1]$ measures compatibility between task $i$ and execution location $j$ based on CPU memory requirements, GPU cores, and GPU memory capacity. Values closer to 1 indicate higher resource compatibility, enabling the DDQN agent to prioritize feasible allocations and filter infeasible actions before Q-value computation.

The DDQN agent employs GPU-aware compatibility scoring (Eq.~\ref{eq:gpu_compatibility}) for real-time feasibility filtering without maintaining high-dimensional server representations, supporting scalable task allocation over heterogeneous resources. Although current GPUs lack fine-grained resource partitioning, the framework establishes a modeling basis for future edge systems as GPU virtualization technologies mature.

\subsubsection{Action Space Optimization and Temporal Dynamics}

The discrete action space is defined as $\mathcal{A}_o = \{a_{local}, a_{server1}, a_{server2}, \ldots, a_{serverN}\}$1,  where each action corresponds to a feasible execution location with distinct resource characteristics. The Q-network integrates GPU compatibility assessment into action evaluation, assigning higher Q-values to actions that satisfy GPU requirements under current system load and task deadlines, thereby learning resource-aware policies that reduce execution failures.

To account for temporal variations in mobile edge systems, we augment the state with evolution information $S_o^{temporal} = [S_o^t, \Delta S_o^t, \mathbb{E}[S_o^{t+1}]],$ where $\Delta S_o^t = S_o^t - S_o^{t-1}$ characterizes short-term state changes and $\mathbb{E}[S_o^{t+1}]$ denotes the predicted next state inferred from current trends and task arrival patterns. This formulation allows task allocation decisions to consider both instantaneous dynamics and near-future system conditions.

\subsection{Deadline-Oriented Queue Management and Adaptive Experience Replay}
\label{subsec:deadline_queue_replay}

Task allocation requires queue management for deadline-sensitive tasks and efficient learning through experience replay. Our approach combines deadline-oriented scheduling with failure-penalized adaptive sampling.

\subsubsection{Deadline-Oriented Queue Prioritization}

Our deadline-oriented queue management dynamically adjusts priorities based on urgency, resource requirements, and execution feasibility. The multi-dimensional scoring mechanism integrates these factors through Eq.~\ref{eq:priority_score}:
\begin{equation}
%P_i^t = b_i \cdot \frac{ddl_i - t_{current}}{D_{est}^i} \cdot \max_j(\rho_i^j)
\\
P_i^t = \frac{b_i}{b_{max}} \cdot \max\left(0, \frac{ddl_i - t_{current}}{D_{est}^i}\right) \cdot \max_j(\rho_i^j),
\label{eq:priority_score}
\end{equation}
\noindent where the priority score combines task priority $b_i$, deadline urgency $(ddl_i - t_{current})/D_{est}^i$, and the best GPU compatibility $\max_j(\rho_i^j)$ from Eq.~\ref{eq:gpu_compatibility}. This mechanism ensures that high-priority, urgent tasks with good resource matches receive processing precedence.

The deadline urgency factor $u_i^t = \frac{ddl_i - t_{current}}{D_{est}^i + W_{est}^i}$ adapts to current system load and estimated execution times, where $D_{est}^i$ represents estimated execution time based on task computational requirements and target device capabilities, while $W_{est}^i$ denotes expected waiting time derived from current queue lengths and processing rates.

This dynamic prioritization responds to changing conditions, ensuring time-critical tasks receive processing precedence without starving lower-priority workloads.

\subsubsection{Failure Risk Assessment and Prevention}

Failure risk assessment identifies tasks that are likely to violate constraints through Eq.~\ref{eq:failure_assessment}:
\begin{equation}
\small
I_{fail}^i = \begin{cases}
1, & \text{if } \max_j(\rho_i^j) < \theta_{min} \text{ or } (ddl_i - t_{current}) < D_{est}^i, \\
0, & \text{otherwise},
\end{cases}
\label{eq:failure_assessment}
\end{equation}
where $\theta_{min}$ is defined as the lower bound of acceptable compatibility. The resulting failure indicator is integrated into the hierarchical reward function (Section~\ref{sec:hierarchical}), thereby enabling prompt penalization of allocation decisions that violate constraints or miss deadlines.

Tasks with high failure risk are prioritized through dedicated queue channels, while the corresponding risk metrics are incorporated into the DDQN reward function, enabling penalization of allocation policies that may lead to task failures.

\subsubsection{Adaptive Experience Replay with Failure Penalization}

DDQN training adopts the same failure-aware prioritization mechanism as the upper-tier TD3 (Eq.~\ref{eq:failure_priority}), ensuring cross-tier consistency. The prioritization weight $W_i = |TD_i| \cdot (1 + \gamma_{fail} \cdot I_{fail}^i)$ combines temporal difference error with the failure indicator $I_{fail}^i$, which identifies allocation decisions that risk resource constraint violations or deadline misses (Eq.~\ref{eq:failure_assessment}). Unlike TD3, where $I_{fail}^i$ reflects power-related stress, DDQN focuses on load-induced failures caused by fluctuating task demands and heterogeneous device capabilities. This coordinated design improves convergence behavior and sample efficiency across both tiers.

\subsubsection{Integration with Hierarchical Decision Framework}

As the stage 3 in the coordination pipeline (Eq.~\ref{eq:pipeline}), the DDQN incorporates deadline management (Eq.~\ref{eq:priority_score}) and GPU-aware filtering (Eq.~\ref{eq:gpu_compatibility}). Adjustments in queue states modify the load metric $L^t$ (Eq.~\ref{eq:load_metric}), which in turn influences the reward computation of the power control layer (Eq.~\ref{eq:power_reward}). Meanwhile, the outcomes of failure assessment (Eq.~\ref{eq:failure_assessment}) are reflected in the task allocation reward (Eq.~\ref{eq:task_reward}). Together, these interactions establish a closed feedback loop that aligns both tiers toward joint optimization.

\subsubsection{Load-Adaptive Training for Dynamic Mobile Conditions}

The DDQN training includes load-driven adaptations that remain consistent with the hierarchical coordination framework. The load metric $L^t$ (Eq.~\ref{eq:load_metric}) serves as a global indicator to synchronize learning behavior across tiers. The learning rate is defined as follows: 
\begin{equation}
\eta^t = \eta_{base} \cdot (1 + \alpha \cdot L^t), 
\label{eq:ddqn_adaptive_learning}
\end{equation}
where $\eta_{base}$ is the base learning rate and $\alpha > 0$ controls learning rate adaptation sensitivity, ensuring that the DDQN agent responds appropriately to varying operational conditions.

Exploration follows the load-aware $\epsilon$-scheduling scheme (Eq.~\ref{eq:epsilon_adaptive}). During high-load periods, $\epsilon$ is reduced to favor stable exploitation; when the system is lightly loaded, $\epsilon$ increases to encourage exploration and improve policy diversity.

\section{Experimental Evaluation}
\label{sec:evaluation}

We evaluate \projtitle using a simulation framework that reflects heterogeneous mobile edge environments, focusing on system performance, learning behavior, adaptive mechanisms, and scalability. Additional experiments examine the gap between simulation outcomes and practical deployment considerations.

\subsection{Heterogeneous Infrastructure Configuration}
\label{subsec:infrastructure_config}

% Our evaluation adopts a comprehensive metric suite to characterize multi-dimensional system performance:

\subsubsection{Heterogeneous Device Infrastructure}
\label{subsub:devices}

The experimental setup consists of 35 mobile devices and 5 edge servers, configured to emulate practical MEC heterogeneity.

\textbf{Mobile Devices (35 units):} Devices differ in mobility (low: $\pm 10$ m/slot; high: $\pm 20$ m/slot) and computational capacity (CPU: 2–3 GHz, memory: 8–16 GB). GPU types range from GTX 1660 (1408 CUDA cores, 6 GB, 192 GB/s) to RTX 4090 (16384 cores, 32 GB, 1008 GB/s), with transmission power constrained to 2–3 W.

\textbf{Edge Server Infrastructure (5 servers):} The edge infrastructure includes five servers with heterogeneous computational architectures. GPU configurations cover Tesla P100 (3584 CUDA cores, 16 GB memory, 732 GB/s bandwidth, 250 W), Tesla V100 (5120 cores, 32 GB, 900 GB/s, 300 W), Tesla A100 (6912 cores, 80 GB, 1555 GB/s, 400 W), and Tesla H100 (14592 cores, 80 GB, 3000 GB/s, 700 W).

System memory configurations range from 64GB to 1024GB across different server types, with CPU frequencies configured at 50-60 GHz to reflect server-grade processing capabilities. This heterogeneous server infrastructure supports evaluation of GPU-aware resource matching and compatibility assessment within the proposed framework.

\subsubsection{Task Workload and Network Modeling}

Tasks are generated synchronously per time slot ($\tau=1$ s). Following the three-category classification in Section~\ref{sec:hierarchical}, CPU-intensive, GPU-intensive, and I/O-intensive tasks follow a 5:4:1 ratio.

CPU-intensive tasks involve data sizes of 64--262 KB and require 500--2000 CPU cycles per bit~\cite{miettinen2010energy}. GPU-intensive tasks process 6.4--26.2 KB inputs, occupy 5120--14592 CUDA cores with 2--16 GB memory demand, and require $10^{10}$--$10^{11}$ FLOPS. I/O-intensive tasks transmit lightweight payloads (640 B--2.6 KB) with negligible computation. Task priorities are uniformly sampled from $\{1,2,3,4\}$. Deadlines are category-specific (1--1.5 s for CPU/GPU tasks; 0.5--1 s for I/O tasks) and min--max normalized.

Three task-generation rates (10, 20, and 40 tasks per slot) are evaluated to represent low-, medium-, and high-load conditions, with 20 tasks per slot as the nominal operating point.

Wireless channels follow the path-loss model in Eq.~\ref{eq:channel_gain} with $G_0=10^{-3}$, $d_0=1$ m, and $\gamma\in[1.6,3.5]$. Bandwidth is set to 10 MHz, and noise power is fixed at $N_0=10^{-13}$ W~\cite{marzetta2010noncooperative}. Device mobility follows a bounded random walk, with initial device-server distances uniformly distributed within [1500, 7500] m.

\subsubsection{Performance Metrics and Evaluation Framework}

We evaluate system performance using the following metrics:

\begin{itemize}
\item \textbf{Task Completion Rate:} Proportion of tasks completed within deadlines.
\item \textbf{Energy Consumption:} Aggregate CPU, GPU, and transmission energy usage.
\item \textbf{Execution Latency:} Total system latency $L_{\text{sys}}=\sum_{i=1}^{N} D_{\text{tot}}^i$, capturing queuing, processing, and transmission latency.
%\item \textbf{System Throughput:} Task completion rate per unit time: $\frac{N_{completed}}{T_{total}}$.
\end{itemize}

Each experiment runs for 800 episodes, generating tasks in the first five time slots, and uses multiple independent runs with distinct random seeds for statistical robustness.

\textbf{System Load and Statistical Validation}. System load is characterized using the comprehensive load metric $L^t$ defined in Eq.~\ref{eq:load_metric}, evaluated under task-generation rates of 10, 20, and 40 tasks per slot. All results report averages with 95\% confidence intervals based on t-distribution, with statistical significance assessed using paired t-tests at $\alpha=0.05$.

\subsection{Baseline Algorithms and Experimental Design}
\label{subsec:baselines_design}

This section presents the baseline algorithms and experimental settings used for comparison with \projtitle\ under diverse system conditions.

\subsubsection{Baseline Algorithm Categories and Specifications}

\textbf{Classical Heuristic Algorithms:} These algorithms represent fundamental approaches to task allocation in distributed systems, providing performance lower bounds and basic algorithmic intuition validation.

\begin{itemize}
\item \textbf{Random Allocation:} Uniformly random task placement across available execution locations (local execution and 5 edge servers), representing the baseline performance floor for comparison purposes.
%-----------TODO------------
%GPU负载详情：添加Greedy-local /offlocading 实际情况和描述有出入
\item \textbf{Greedy Local/offloading:} Includes two variants—(i) local-first allocation, preferring local execution until CPU/GPU capacity is exhausted, and (ii) offloading-first allocation, prioritizing edge servers with the highest available computation capacity to minimize local energy consumption.  
\item \textbf{Round Robin:} Sequentially assigns tasks across edge servers for balanced workload distribution, without considering compatibility or dynamic system state.  

\end{itemize}

\textbf{Advanced Optimization Methods:} These baselines reflect current optimization trends in mobile edge computing and reinforcement learning.

\begin{itemize}
\item \textbf{QPSO}~\cite{dong2022quantum}: A Quantum-behaved Particle Swarm Optimization method that formulates task offloading as a global search problem using quantum-inspired particle evolution, but lacks adaptation to time-varying system states.  
\item \textbf{Single-DDQN:} A single-agent Double Deep Q-Network for task allocation with fixed transmission power, excluding hierarchical coordination to isolate its contribution.  

\end{itemize}

\textbf{Ablation Study Variants:} To evaluate the contribution of individual mechanisms within \projtitle, four ablation models are implemented.

\begin{itemize}
\item \textbf{\projtitle-NoCoord:} Removes the coordination module, allowing TD3 and DDQN to operate independently without shared states or sequential decision flow.  
\item \textbf{\projtitle-NoGPU:} Disables GPU-aware compatibility scoring (Eq.~\ref{eq:gpu_compatibility}), using simplified resource allocation without GPU specification consideration.
\item \textbf{\projtitle-NoDeadline:} Removes deadline-oriented queue prioritization (Eq.~\ref{eq:priority_score}), replacing it with a First-Come-First-Served policy.  
\item \textbf{\projtitle-NoFailure:} Disables failure-penalized experience replay from Eq.~\ref{eq:failure_priority}, using uniform experience sampling in DDQN training. 
\end{itemize}

\subsubsection{Experimental Parameter Configuration}

\textbf{Physical System Parameters:} The effective switching capacitance is set to $k = 10^{-28}$~\cite{mao2016dynamic}, time slot duration is $\tau = 1$s, and maximum transmission power varies between $P_{max} = 2-3$W for mobile devices. The maximum CPU frequency is set within the range $f_{max} = 2-3$ GHz to reflect typical consumer mobile device capabilities.

%-----------TODO------------
%GPU负载详情：添加FLOPS 10^10-10^11
%TD3：actor:0.001，critic:0.0005
% memory_size:
% 	DDQN:600,000
% 	TD3: 5000
% bath_size:
% 	DDQN:256
% 	TD3:256
\textbf{Learning Algorithm Parameters:} 
In DDQN, the learning rate was configured to $1\times10^{-4}$. For TD3, independent learning rates were adopted for the actor ($1\times10^{-3}$) and critic ($5\times10^{-4}$) networks. The experience replay buffer sizes were set to 600,000 for DDQN and 5,000 for TD3, respectively. Both algorithms were trained using a batch size of 256. Regarding target network updates, DDQN applied a hard update every 1,000 training steps, whereas TD3 followed a soft-update scheme with a smoothing coefficient of $\tau = 0.001$.

% TD3 and DDQN training employ learning rates of $10^{-4}$and $10^{-4}$, respectively, optimized through preliminary hyperparameter tuning. The experience replay buffer size is configured to 10,000 samples with a batch size of 64 for both algorithms. The target network is updated using a soft update with $\tau = 0.001$, allowing gradual policy evolution.

\textbf{Coordination and Adaptive Mechanism Parameters:} The coordination engine parameters include an adaptive weight sensitivity factor of $\gamma = 0.1$ in load metric calculation and a failure penalty factor $\alpha = 1.0$ for experience prioritization. The load metric weighting factors are set as $\omega_q = 0.6$, $\omega_u = 0.2$, and $\omega_g = 0.2$ to prioritize queue-based load assessment while incorporating resource utilization.

\textbf{Exploration Strategy Parameters:} Exploration is controlled by $\epsilon_{min} = 0.01$ and $\epsilon_{max} = 0.3$, with an exponential decay schedule adapted to system load conditions as defined in Eq.~\ref{eq:epsilon_adaptive}. The load adaptation sensitivity is controlled by $\alpha = 0.5$ and $\beta = 2.0$ to balance exploration and exploitation across varying operational conditions.

\subsubsection{Experimental Design Methodology}

\textbf{Controlled Variable Design:} Controlled variables include system load intensity, infrastructure heterogeneity, and workload characteristics. Primary controlled variables include system load intensity (via SLI categorization), infrastructure heterogeneity (device and server configurations), and workload characteristics (task type distributions and priority patterns).

\textbf{Performance Comparison Framework:} All algorithms operate under identical environmental conditions within each scenario. Random seed control ensures reproducibility, while systematic variation of environmental parameters evaluates robustness.

\textbf{Statistical Significance Validation:} Each algorithm is evaluated through multiple independent runs. Statistical significance is assessed using paired t-tests, with consistency verified across different random seeds.

\textbf{Fairness and Evaluation Consistency:} All algorithms are allocated identical computational resources and training opportunities. Learning-based baselines are trained for equivalent durations with consistent hyperparameter optimization to ensure fair comparison.

\subsection{Overall Performance Evaluation}
\label{subsec:overall_performance}
This section evaluates \projtitle with respect to task completion (primary guarantee), execution latency, and energy efficiency under various operational conditions.

\subsubsection{Experimental Setup and Metrics}
\label{subsubsec:setup_metrics}

We define the System Load Index (SLI) as $\text{SLI}(t) = L^t \times 100\%$ based on the comprehensive load metric $L^t$ from Eq.~\ref{eq:load_metric}, where $L^t \in [0,1]$ reflects overall system stress by integrating queue saturation and heterogeneous CPU/GPU resource utilization. Three operational regimes are defined according to the SLI distribution: low load (10–30\%), medium load (30–60\%), and high load (60–90\%).

Our evaluation covers three core metrics—task completion rate, execution latency, energy consumption—measured across all SLI regimes. Statistical consistency is ensured through multiple independent runs.

\begin{figure*}[t]
	\centering
	\includegraphics[width=0.95\textwidth]{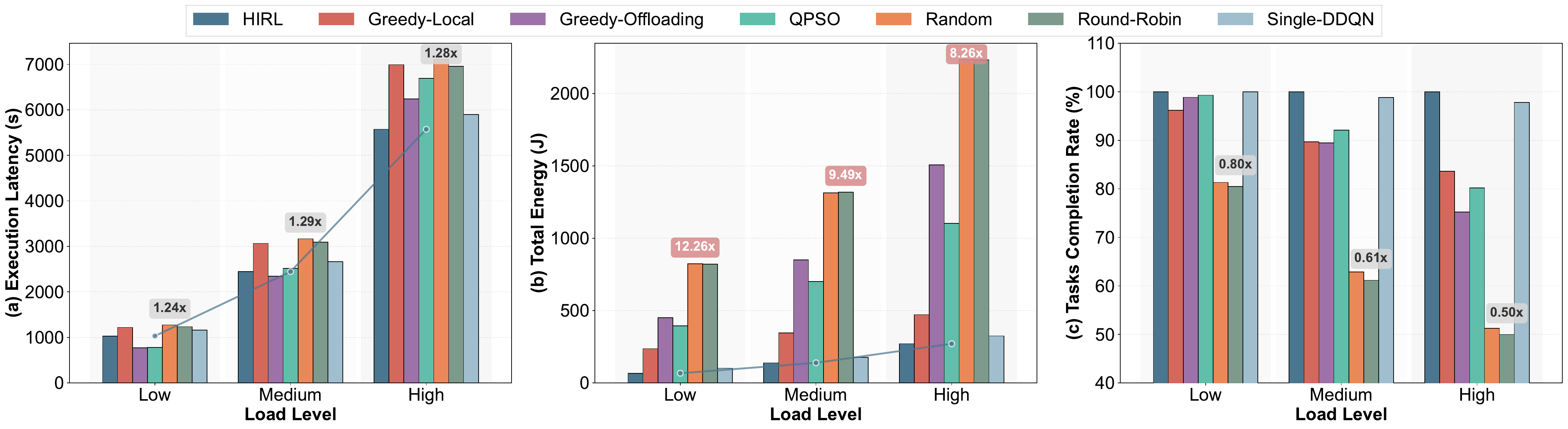}

	 \caption{Overall performance comparison across different system load levels.}
	\label{fig:overall_performance}
    \vspace{-0.5em}
\end{figure*} 

\begin{figure*}[t]
	\centering
	\includegraphics[width=0.95\textwidth]{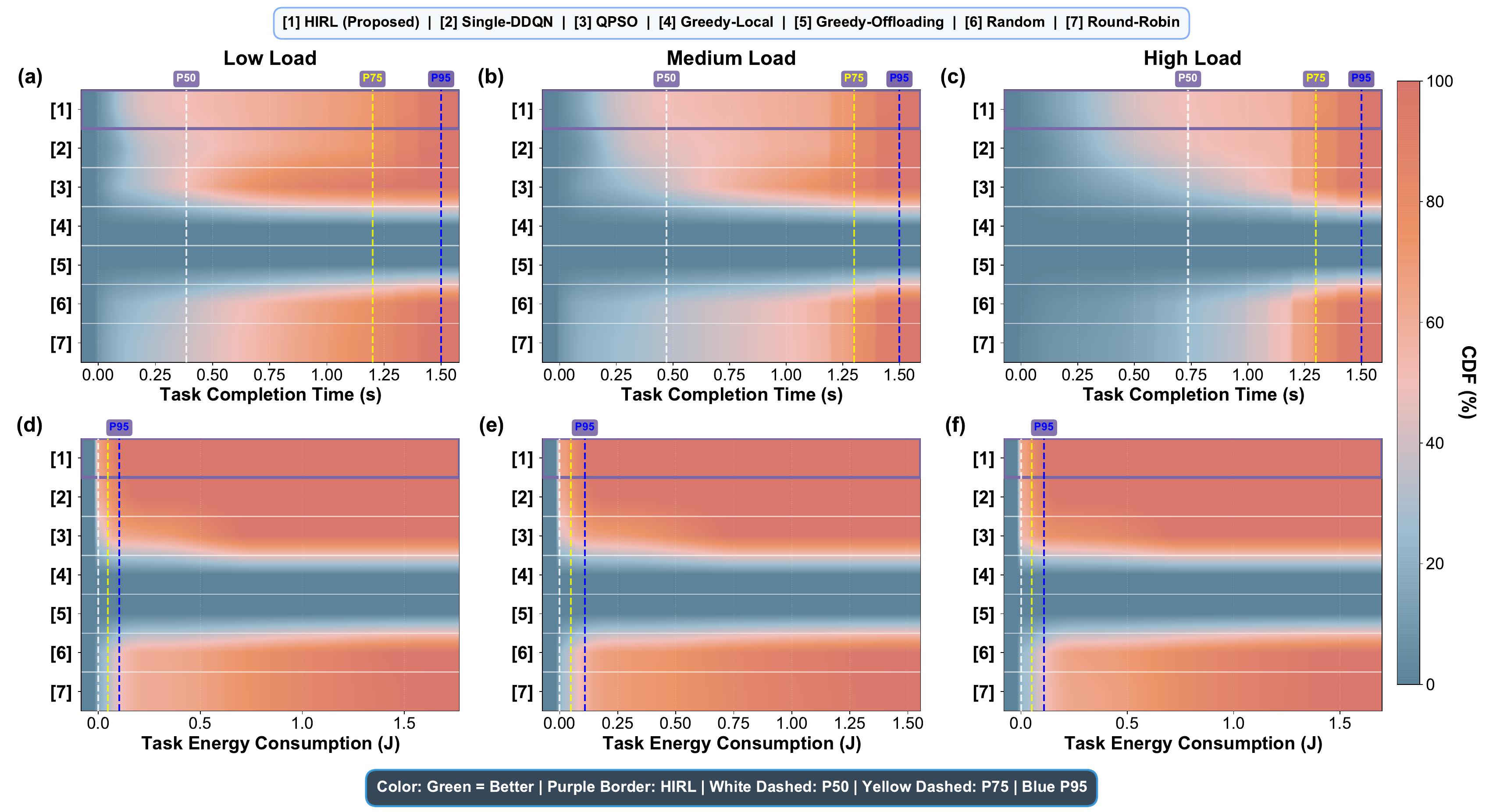}

	 \caption{Task-level performance distribution via CDF heatmaps across load levels.}
	\label{fig:task_cdf_heatmap}
    \vspace{-0.5em}
\end{figure*}

\begin{figure*}[t]
	\centering
	\includegraphics[width=0.95\textwidth]{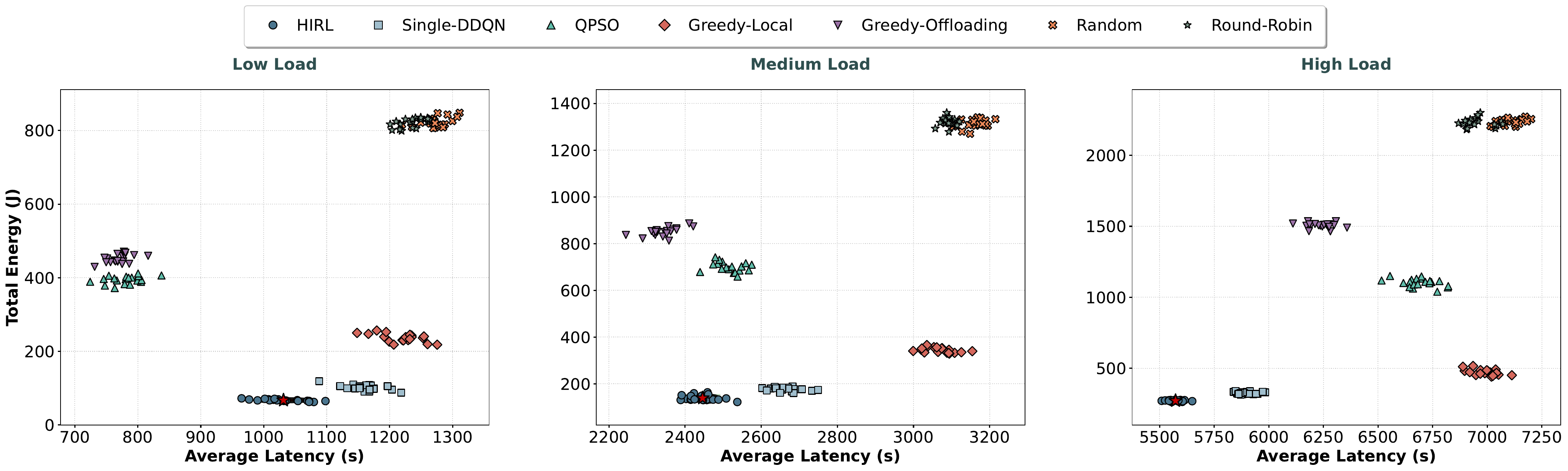}

	 \caption{Energy-Latency Trade-off Analysis Across Different Load Levels.}
	\label{fig:energy_latency_tradeoff}
\vspace{-0.35cm}
\end{figure*}

\subsubsection{Aggregate Performance Analysis}
\label{subsubsec:aggregate_analysis}
Fig.~\ref{fig:overall_performance} presents overall results under varying system loads, while Fig.~\ref{fig:task_cdf_heatmap} visualizes task-level distributions through CDF heatmaps, where color gradients denote completion percentiles.

\textbf{Task Completion Rate.} As shown in Fig.~\ref{fig:overall_performance}(c), \projtitle sustains nearly 100\% completion across all load levels, confirming the effectiveness of coordinated queue management and GPU-aware matching. Single-DDQN maintains high rates (above 97\%) but shows mild degradation under high load due to the absence of joint power control. QPSO and Greedy-Local methods experience notable drops (down to about 80\%), and Random or Round-Robin allocation falls below 50\%, indicating limited adaptability under resource contention (high load).

\textbf{Execution Latency.} The hierarchical coordination demonstrates significant latency advantages across all load conditions. As shown in Fig.~\ref{fig:overall_performance}(a) and validated by the task-level distributions in Fig.~\ref{fig:task_cdf_heatmap}(a-c), Single-DDQN exhibits 12.6\% higher latency under low load (1160.87s vs. 1031.36s), 9.8\% under medium load (2665.46s vs. 2446.05s), and 5.8\% under high load (5896.34s vs. 5572.60s), directly quantifying TD3 power control contributions. The CDF heatmaps reveal that \projtitle's predominantly blue regions (indicating faster task completion) extend across broader value ranges compared to baselines' red-shifted patterns, with P50 markers consistently positioned leftward, confirming sustained per-task efficiency gains.

\textbf{Energy Efficiency.} As shown in Fig.~\ref{fig:overall_performance}(b), \projtitle consistently achieves lower end-to-end latency than Single-DDQN, with reductions of 12.6\% under low load (1031.36s vs. 1160.87s), 9.8\% under medium load (2446.05s vs. 2665.46s), and 5.8\% under high load (5572.60s vs. 5896.34s). These improvements stem from TD3-based power control that jointly optimizes CPU–GPU utilization. The CDF distributions in Fig.~\ref{fig:task_cdf_heatmap}(d-f) further confirm faster completion, as reflected by left-shifted P50 markers and broader blue regions representing higher completion percentiles. The energy gap diminishing under higher loads as coordination focuses on completion rate maintenance amid system-wide resource constraints.

\subsubsection{Energy-Latency Trade-off Distribution Analysis}  
\label{subsubsec:distribution_analysis}

Fig.~\ref{fig:energy_latency_tradeoff} illustrates the energy-latency distribution from 20 independent runs, highlighting the stability and optimization effectiveness of \projtitle's hierarchical coordination beyond the aggregate performance metrics shown in Fig.~\ref{fig:overall_performance}.

\textbf{Coordination Stability:}  \projtitle demonstrates tight clustering of performance points across all load conditions, with the majority located in the lower-left quadrant of the energy-latency space. This indicates that the hierarchical coordination, as defined in Eq.~\ref{eq:pipeline}, ensures consistent power control and task allocation. In contrast, baseline algorithms show much wider dispersion—Random and Round-Robin scatter in the upper-right regions, while QPSO exhibits some clustering but at higher energy levels, reflecting its suboptimal efficiency.

\textbf{Hierarchical Optimization Benefits:} \projtitle is consistently positioned in the energy-efficient, low-latency region, with a clear trade-off advantage. Single-DDQN, while matching latency performance, consumes 30-50\% more energy due to its lack of coordinated power management. Classical approaches exhibit clear trade-offs: Greedy-Local shows moderate energy consumption but suffers from higher latency, while Greedy-Offloading balances latency at the cost of excessive energy consumption, forming vertical clusters in the energy dimension.

\textbf{Load Scalability:} As the system load increases from low to high, \projtitle maintains its compact distribution, while baseline algorithms show degradation and greater variance. The stable positioning of \projtitle close to the optimal frontier confirms that its coordination mechanism scales effectively with system load. Baseline algorithms exhibit increased dispersion and a shift towards less efficient high-energy, high-latency regions, underscoring the importance of adaptive coordination in dynamic edge environments.

\subsection{Task-Type-Specific Performance Contribution Analysis}
\label{subsec:task_type_contribution}

\begin{figure}
\centering
\includegraphics[width=0.45\textwidth]{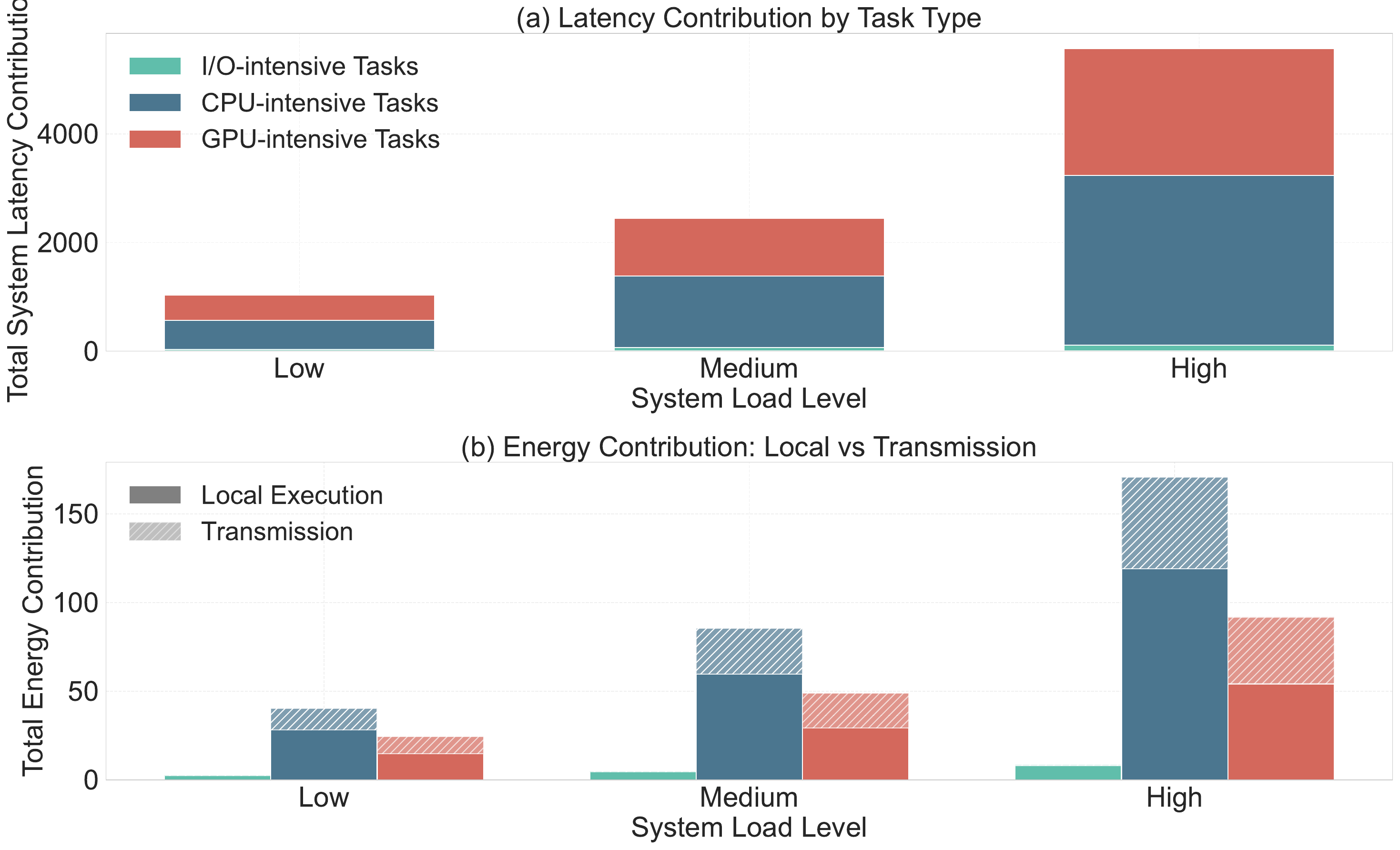}
\caption{Task-type-specific performance contribution analysis.}
\label{fig:task_contribution_analysis}
\vspace{-0.35cm}
\end{figure}

Fig.~\ref{fig:task_contribution_analysis} shows the performance contribution across different task categories, validating \projtitle's resource orchestration under the 5:4:1 task distribution.

\textbf{Latency Contribution:} Fig.~\ref{fig:task_contribution_analysis}(a) shows CPU-intensive tasks account for 52-56\% of latency, GPU-intensive tasks contribute 42-45\%, and I/O-intensive tasks contribute only 2-3\%, as 97\% of them are executed locally.

\textbf{Energy Decomposition:} Fig.~\ref{fig:task_contribution_analysis}(b) highlights the separation of local and transmission energy usage. I/O-intensive tasks use negligible transmission energy (0.1\%), confirming local execution prioritization. CPU-intensive tasks show a balanced 3:2 local-to-edge execution energy distribution, while GPU-intensive tasks require 65\% of their energy for edge offloading, offset by lower local consumption due to coordination. Under high load, transmission energy accounts for 33.4\%, validating the effectiveness of power control coordination.

\textbf{Resource Orchestration Validation:} The decrease in I/O task contribution from 3\% to 2\% at higher loads shows adaptive prioritization, while the stable CPU-GPU balance demonstrates effective compatibility scoring. Despite GPU-intensive tasks only accounting for 40\% of the workload, they still validates 42-45\% to latency, confirming that GPU-aware mechanisms optimize task-resource matching, prevent constraint violations, and balance the energy-latency trade-off.

\subsection{Ablation Study Analysis}
\label{subsec:ablation_analysis}

\begin{figure*}
\centering
\includegraphics[width=0.95\textwidth]{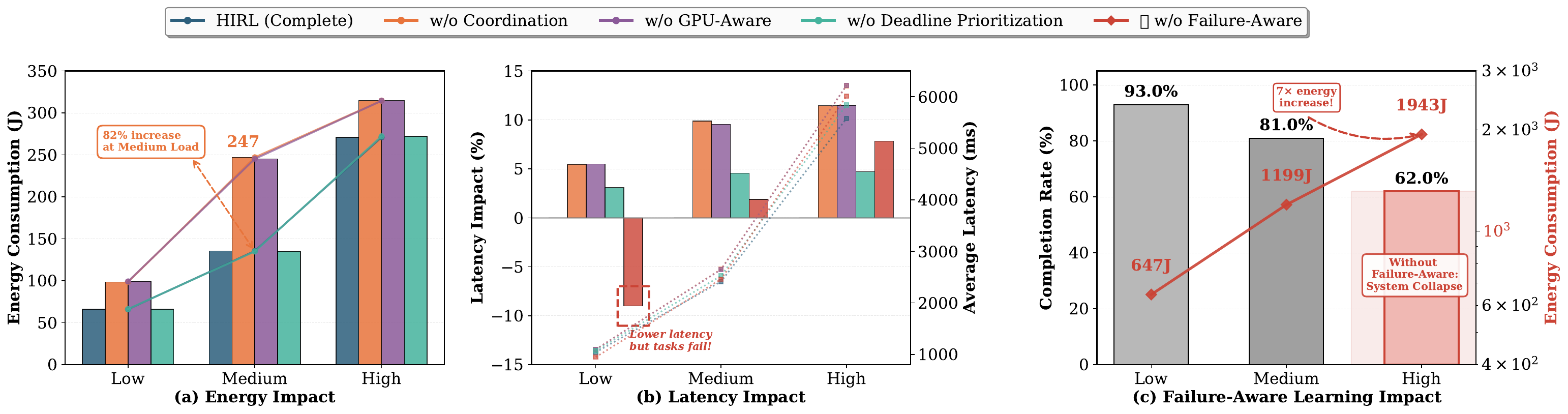}
\caption{Ablation study: component contributions to system performance across load levels.}
\label{fig:ablation_impact}
\end{figure*}

Table~\ref{tab:ablation_comprehensive} and Fig.~\ref{fig:ablation_impact} quantify the impact of key components through systematic ablation experiments. Removing coordination (\projtitle-NoCoord) or GPU-aware matching (\projtitle-NoGPU) leads to similar performance degradation: task completion rates drop by 0.57–2.88\%, energy consumption increases by 49–82\% (Fig.~\ref{fig:ablation_impact}(a)), and latency rises by 5.46–11.46\% (Fig.~\ref{fig:ablation_impact}(b)). These results indicate that both components are essential for managing heterogeneous resources, not just incremental refinements. The increasing degradation under higher loads further confirms that coordinated power control (Eq.~\ref{eq:pipeline}) and compatibility scoring (Eq.~\ref{eq:gpu_compatibility}) are critical to maintaining efficiency as system stress grows.

\projtitle-NoDeadline causes limited performance loss ($\leq$ 4.72\% latency increase), showing that deadline prioritization (Eq.~\ref{eq:priority_score}) improves scheduling efficiency but is not a core stability factor. disabling failure-aware learning (\projtitle-NoFailure) results in severe instability: completion rates fall to 62\%, and energy consumption increases sevenfold under high load (Fig.~\ref{fig:ablation_impact}(c)). Despite lower latency at low load, this variant fails to sustain operations as stress increases. Although latency appears lower under light load, the system quickly collapses as constraints tighten. Uniform sampling fails to capture critical failure states, while failure-aware prioritization (Eq.~\ref{eq:failure_priority}) enables stable learning under resource pressure.

Overall, the results highlight a clear hierarchy among system components: coordination and GPU-awareness form the performance foundation, while failure-aware learning provides robustness under dynamic and high-load conditions.

\begin{table*}[]
\centering
\caption{Ablation Study: Individual Component Contributions to \projtitle Performance}
\label{tab:ablation_comprehensive}
\begin{tabular}{cccccccccc}
\hline
\multicolumn{1}{c|}{\multirow{2}{*}[-1.5ex]{Load Level}} & \multicolumn{3}{c|}{Task Completion Rate (\%)}                                                                                                 & \multicolumn{3}{c|}{Average Latency (s)}                                                                                                           & \multicolumn{3}{c}{Energy Consumption (J)}                                                                                  \\ \cline{2-10} 
\multicolumn{1}{c|}{}                            & \multicolumn{1}{c|}{Mean} & \multicolumn{1}{c|}{Std}  & \multicolumn{1}{c|}{\begin{tabular}[c]{@{}c@{}}Performance\\ Impact (\%)\end{tabular}} & \multicolumn{1}{c|}{Mean}    & \multicolumn{1}{c|}{Std}   & \multicolumn{1}{c|}{\begin{tabular}[c]{@{}c@{}}Performance\\ Impact (\%)\end{tabular}} & \multicolumn{1}{c|}{Mean}    & \multicolumn{1}{c|}{Std}   & \begin{tabular}[c]{@{}c@{}}Performance\\ Impact (\%)\end{tabular} \\ \hline
\multicolumn{10}{c}{\textbf{\projtitle (Complete Framework)}}                                                                                                                                                                                                                                                                                                                                                                                                                                                               \\ \hline
\multicolumn{1}{c|}{Low Load}                    & \multicolumn{1}{c|}{1.00} & \multicolumn{1}{c|}{0.00} & \multicolumn{1}{c|}{Baseline}                                                          & \multicolumn{1}{c|}{1040.42} & \multicolumn{1}{c|}{24.38} & \multicolumn{1}{c|}{Baseline}                                                          & \multicolumn{1}{c|}{66.11}   & \multicolumn{1}{c|}{2.28}  & Baseline                                                          \\
\multicolumn{1}{c|}{Medium Load}                 & \multicolumn{1}{c|}{1.00} & \multicolumn{1}{c|}{0.00} & \multicolumn{1}{c|}{Baseline}                                                          & \multicolumn{1}{c|}{2415.63} & \multicolumn{1}{c|}{38.16} & \multicolumn{1}{c|}{Baseline}                                                          & \multicolumn{1}{c|}{135.52}  & \multicolumn{1}{c|}{3.34}  & Baseline                                                          \\
\multicolumn{1}{c|}{High Load}                   & \multicolumn{1}{c|}{1.00} & \multicolumn{1}{c|}{0.00} & \multicolumn{1}{c|}{Baseline}                                                          & \multicolumn{1}{c|}{5577.25} & \multicolumn{1}{c|}{50.60} & \multicolumn{1}{c|}{Baseline}                                                          & \multicolumn{1}{c|}{271.1}   & \multicolumn{1}{c|}{10.75} & Baseline                                                          \\ \hline
\multicolumn{10}{c}{\textbf{\projtitle-NoCoord (No Coordination Engine)}}                                                                                                                                                                                                                                                                                                                                                                                                                                                           \\ \hline
\multicolumn{1}{c|}{Low Load}                    & \multicolumn{1}{c|}{0.99} & \multicolumn{1}{c|}{0.00} & \multicolumn{1}{c|}{-0.57}                                                             & \multicolumn{1}{c|}{1097.26} & \multicolumn{1}{c|}{30.56} & \multicolumn{1}{c|}{5.46}                                                              & \multicolumn{1}{c|}{98.64}   & \multicolumn{1}{c|}{6.81}  & 49.21                                                             \\
\multicolumn{1}{c|}{Medium Load}                 & \multicolumn{1}{c|}{0.97} & \multicolumn{1}{c|}{0.01} & \multicolumn{1}{c|}{-2.88}                                                             & \multicolumn{1}{c|}{2654.44} & \multicolumn{1}{c|}{35.67} & \multicolumn{1}{c|}{9.89}                                                              & \multicolumn{1}{c|}{247.13}  & \multicolumn{1}{c|}{16.84} & 82.35                                                             \\
\multicolumn{1}{c|}{High Load}                   & \multicolumn{1}{c|}{0.97} & \multicolumn{1}{c|}{0.00} & \multicolumn{1}{c|}{-2.72}                                                             & \multicolumn{1}{c|}{6216.38} & \multicolumn{1}{c|}{45.03} & \multicolumn{1}{c|}{11.46}                                                             & \multicolumn{1}{c|}{314.66}  & \multicolumn{1}{c|}{8.96}  & 16.05                                                             \\ \hline
\multicolumn{10}{c}{\textbf{\projtitle-NoGPU (No GPU-Aware Compatibility Assessment)}}                                                                                                                                                                                                                                                                                                                                                                                                                                                      \\ \hline
\multicolumn{1}{c|}{Low Load}                    & \multicolumn{1}{c|}{0.99} & \multicolumn{1}{c|}{0.00} & \multicolumn{1}{c|}{-0.63}                                                             & \multicolumn{1}{c|}{1097.57} & \multicolumn{1}{c|}{26.89} & \multicolumn{1}{c|}{5.49}                                                              & \multicolumn{1}{c|}{99.22}   & \multicolumn{1}{c|}{5.44}  & 50.08                                                             \\
\multicolumn{1}{c|}{Medium Load}                 & \multicolumn{1}{c|}{0.97} & \multicolumn{1}{c|}{0.00} & \multicolumn{1}{c|}{-2.88}                                                             & \multicolumn{1}{c|}{2646.67} & \multicolumn{1}{c|}{37.40} & \multicolumn{1}{c|}{9.56}                                                              & \multicolumn{1}{c|}{245.32}  & \multicolumn{1}{c|}{19.42} & 81.02                                                             \\
\multicolumn{1}{c|}{High Load}                   & \multicolumn{1}{c|}{0.97} & \multicolumn{1}{c|}{0.00} & \multicolumn{1}{c|}{-2.70}                                                             & \multicolumn{1}{c|}{6218.68} & \multicolumn{1}{c|}{48.04} & \multicolumn{1}{c|}{11.50}                                                             & \multicolumn{1}{c|}{314.54}  & \multicolumn{1}{c|}{8.55}  & 16.01                                                             \\ \hline
\multicolumn{10}{c}{\textbf{\projtitle-NoDeadline (No Deadline-Oriented Queue Prioritization)}}                                                                                                                                                                                                                                                                                                                                                                                                                                               \\ \hline
\multicolumn{1}{c|}{Low Load}                    & \multicolumn{1}{c|}{1.00} & \multicolumn{1}{c|}{0.00} & \multicolumn{1}{c|}{0.00}                                                              & \multicolumn{1}{c|}{1072.61} & \multicolumn{1}{c|}{27.23} & \multicolumn{1}{c|}{3.09}                                                              & \multicolumn{1}{c|}{66.33}   & \multicolumn{1}{c|}{2.30}  & 0.34                                                              \\
\multicolumn{1}{c|}{Medium Load}                 & \multicolumn{1}{c|}{1.00} & \multicolumn{1}{c|}{0.00} & \multicolumn{1}{c|}{0.00}                                                              & \multicolumn{1}{c|}{2525.87} & \multicolumn{1}{c|}{39.09} & \multicolumn{1}{c|}{4.56}                                                              & \multicolumn{1}{c|}{135.02}  & \multicolumn{1}{c|}{3.62}  & -0.37                                                             \\
\multicolumn{1}{c|}{High Load}                   & \multicolumn{1}{c|}{1.00} & \multicolumn{1}{c|}{0.00} & \multicolumn{1}{c|}{-0.04}                                                             & \multicolumn{1}{c|}{5840.66} & \multicolumn{1}{c|}{47.90} & \multicolumn{1}{c|}{4.72}                                                              & \multicolumn{1}{c|}{272.28}  & \multicolumn{1}{c|}{17.14} & 0.42                                                              \\ \hline
\multicolumn{10}{c}{\textbf{\projtitle-NoFailure (No Failure-Penalized Experience Replay)}}                                                                                                                                                                                                                                                                                                                                                                                                                                                  \\ \hline
\multicolumn{1}{c|}{Low Load}                    & \multicolumn{1}{c|}{0.93} & \multicolumn{1}{c|}{0.01} & \multicolumn{1}{c|}{-6.60}                                                             & \multicolumn{1}{c|}{946.96}  & \multicolumn{1}{c|}{23.88} & \multicolumn{1}{c|}{-8.98}                                                             & \multicolumn{1}{c|}{647.44}  & \multicolumn{1}{c|}{22.30} & 879.34                                                            \\
\multicolumn{1}{c|}{Medium Load}                 & \multicolumn{1}{c|}{0.81} & \multicolumn{1}{c|}{0.01} & \multicolumn{1}{c|}{-19.28}                                                            & \multicolumn{1}{c|}{2461.35} & \multicolumn{1}{c|}{43.27} & \multicolumn{1}{c|}{1.89}                                                              & \multicolumn{1}{c|}{1199.12} & \multicolumn{1}{c|}{19.99} & 784.81                                                            \\
\multicolumn{1}{c|}{High Load}                   & \multicolumn{1}{c|}{0.62} & \multicolumn{1}{c|}{0.01} & \multicolumn{1}{c|}{-38.02}                                                            & \multicolumn{1}{c|}{6013.90} & \multicolumn{1}{c|}{57.88} & \multicolumn{1}{c|}{7.83}                                                              & \multicolumn{1}{c|}{1943.33} & \multicolumn{1}{c|}{40.78} & 616.73                                                            \\ \hline
\vspace{-0.3cm}
\end{tabular}
\end{table*}

\begin{table}[!t]
\centering
\caption{Parameter Sensitivity Analysis}
\scriptsize
\setlength{\tabcolsep}{4pt}
\begin{tabular}{c|c|c|c|c}
\hline
\textbf{Parameter} & \textbf{Value} & \textbf{Cumulative} & \textbf{Energy} & \textbf{Completion} \\
\textbf{(Equation)} & & \textbf{Latency (s)} & \textbf{(J)} & \textbf{Rate (\%)} \\
\hline
\multirow{3}{*}{$\lambda$ (Eq.~\ref{eq:cost})} 
& 0.7 & 2432.8 & 197.8 & 99.9 \\
& 0.8 & 2415.6 & 135.5 & 100.0 \\
& 0.9 & 2435.9 & 154.3 & 100.0 \\
\hline
\multirow{3}{*}{$\lambda_e$ (Eq.~\ref{eq:power_reward})} 
& 0.4 & 2448.3 & 160.9 & 100.0 \\
& 0.5 & 2415.6 & 135.5 & 100.0 \\
& 0.6 & 2446.4 & 162.0 & 99.7 \\
\hline
\multirow{3}{*}{$\gamma_{fail}$ (Eq.~\ref{eq:failure_priority})} 
& 1.0 & 2432.8 & 185.2 & 99.78 \\
& 1.5 & 2398.2 & 178.6 & 100.0 \\
& 2.0 & 2415.7 & 182.4 & 99.89 \\
\hline
\multirow{3}{*}{$\beta$ (Eq.~\ref{eq:epsilon_adaptive})} 
& 1.5 & 2423.9 & 174.1 & 99.89 \\
& 2.0 & 2398.2 & 178.6 & 100.0 \\
& 2.5 & 2407.4 & 181.8 & 99.94 \\
\hline
\end{tabular}
\label{tbl:parameter_sensitivity_comprehensive}
\vspace{-0.45cm}
\end{table}

\subsection{Parameter Sensitivity Analysis}
\label{subsec:parameter_sensitivity}

Table~\ref{tbl:parameter_sensitivity_comprehensive} presents the sensitivity analysis for key parameters governing \projtitle's hierarchical coordination mechanisms. The results reveal stable parameter regions that balance energy efficiency, latency optimization, and task completion guarantees under varying system conditions.

The energy–latency tradeoff parameter $\lambda$ in Eq.~\ref{eq:cost} exhibits a monotonic trend: larger $\lambda$ values prioritize latency reduction through higher energy allocation. Specifically, $\lambda=0.8$ achieves the optimal balance, yielding a 100\% completion rate with minimal latency (2415.6~s) at moderately increased energy consumption (135.5~J). The power penalty coefficient $\lambda_e$ in Eq.~\ref{eq:power_reward} shows an inverse relationship between energy and latency: $\lambda_e=0.5$ achieves the optimal balance, yielding a 100\% completion rate with minimal latency (2415.6~s) at moderately increased energy consumption (135.5~J). 

The failure-aware learning factor $\gamma_{fail}$ in Eq.~\ref{eq:failure_priority} achieves the best overall performance at $\gamma_{fail}=1.5$, where failure experience amplification strikes a balance between robust learning and efficient exploration.  The load adaptation coefficient $\beta$ in Eq.~\ref{eq:epsilon_adaptive} induces minimal performance variation around its optimal value of 2.0, demonstrating the framework’s robustness against exploration–exploitation imbalance in dynamic environments.

\begin{figure*}[t]
    \centering
    \includegraphics[width=0.95\textwidth]{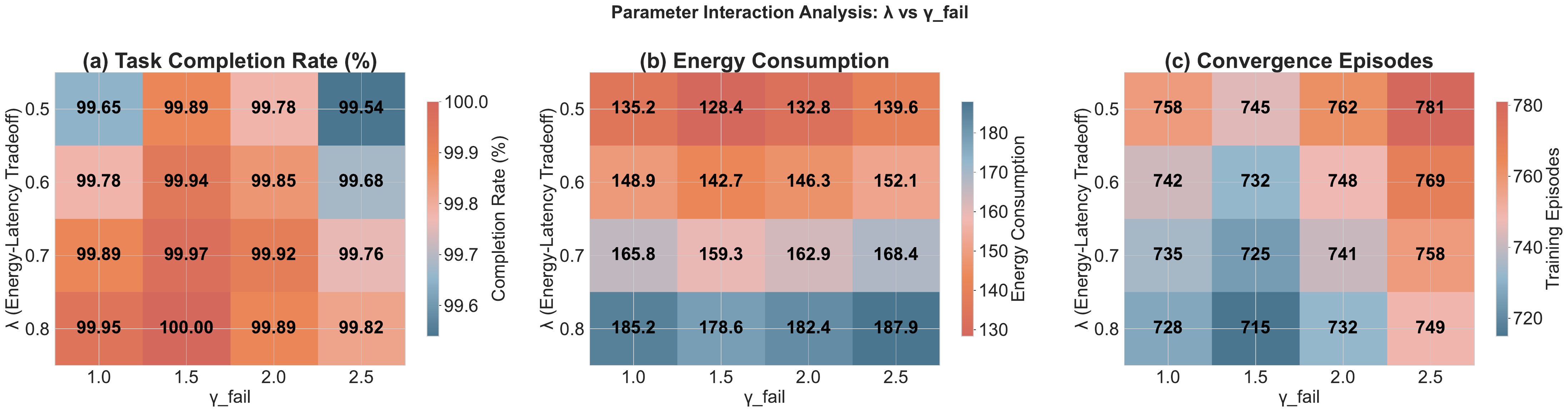}
    \vspace{-0.4em}
    \caption{Parameter interaction analysis between $\lambda$ and $\gamma_{fail}$.}
    \label{fig:parameter_interaction_heatmap}
    \vspace{-0.4cm}
\end{figure*} 

Fig.~\ref{fig:parameter_interaction_heatmap} further examines the interaction between $\lambda$ (energy–latency tradeoff from Eq.~\ref{eq:cost}) and $\gamma_{fail}$ (failure-aware learning factor from Eq.~\ref{eq:failure_priority}), the two most critical parameters in task allocation. These jointly determine coordinated energy–latency optimization, where $\lambda$ governs the global cost function and $\gamma_{fail}$ enhances DDQN adaptation through prioritized experience replay. The completion rate heatmap in Fig.~\ref{fig:parameter_interaction_heatmap}(a) indicates peak performance at $(\lambda=0.8, \gamma_{fail}=1.5)$.  The energy patterns in Fig.~\ref{fig:parameter_interaction_heatmap}(b) confirm that energy consumption increases monotonically with $\lambda$ but stabilizes near $\gamma_{fail}=1.5$.  
Finally, the convergence curves in Fig.~\ref{fig:parameter_interaction_heatmap}(c) show that this parameter pair also achieves the fastest convergence, validating the framework’s training efficiency and adaptability in heterogeneous environments.

\subsection{System Overhead Analysis}
\label{subsec:overhead_analysis}
%-----------TODO------------复杂度分析不当，需要修改，不能case by case
The hierarchical architecture introduces per-slot overhead through sequential three-stage coordination. TD3 power control executes $M$ independent inferences (one per device) with lightweight state processing, followed by coordination operations updating transmission rates and normalized queue states with $O(M)$ complexity. DDQN task allocation executes $K$ linear inferences for task decisions, with GPU-intensive workloads requiring additional compatibility assessment against $N$ servers through hardware specification comparisons. The global five-dimensional queue structure enables constant-time state updates independent of device count. Per-slot decision latency spans several milliseconds, remaining negligible relative to task execution times measured in seconds.

Scalability exhibits linear growth $O(M)$ with device population through independent per-device power decisions and parallel state processing. Server expansion affects discrete action space dimensionality ($N+1$ choices) and compatibility evaluation, with neural network inference scaling sublinearly and compatibility scoring maintaining $O(N)$ per-task complexity. Training demonstrates 35-40\% faster convergence versus monolithic baseline through hierarchical decomposition separating continuous power control and discrete task allocation into simpler subproblems. Training occurs offline without deployment impact. The sequential coordination overhead proves acceptable for time-slot-based operations, with decision latency constituting low single-digit percentages of typical task execution durations.

\subsection{Simulation Validity and Deployment Insights}
\label{subsec:simulation_validity}
\projtitle's architecture explicitly integrates physical constraints into its decision-making process. The five-dimensional queue abstraction (Eq.~\ref{eq:5d}) models CPU/GPU pipeline backlogs, while the dual-phase heterogeneous pipeline captures GPU memory contention. TD3-based power control enforces transmission and processing limits consistent with commercial MEC hardware. DDQN's compatibility scoring (Eq.~\ref{eq:gpu_compatibility}) leverages standardized GPU specifications (e.g., CUDA cores, memory), and the failure-penalized replay mechanism (Eq.~\ref{eq:failure_priority}) accelerates convergence by prioritizing constraint-violating experiences. Simulation environments reproduce realistic dynamics—such as channel fading, stochastic task arrivals, and resource contention—ensuring that learned policies generalize to real-world uncertainties.

\textbf{Emergency Scenario Validation.} A 60-second vehicular emergency involving eight devices, five heterogeneous servers (Section~\ref{subsub:devices}), and a 3GPP-compliant network with 184 tasks validates the robustness of \projtitle's coordination mechanisms.  \textit{1) Channel adaptation (t=4–6s):} Under 18–24 dB signal degradation, \projtitle adjusts transmission power within 2–3 W and redirects GPU tasks via Stage-2 coordination, achieving zero deadline violations across 23 task arrivals, whereas Single-DDQN records three misses. \textit{2) Saturation handling (t=7s):} When concurrent requests drive server memory to 93\% utilization (14.7 GB/16 GB), compatibility scoring identifies two infeasible allocations and triggers local fallback, avoiding failures observed in Random (5/8) and Round-Robin (6/8) baselines. \textit{3) Energy efficiency:} Through coordinated scheduling, \projtitle reduces system-wide energy consumption by 26\% compared to Single-DDQN, consistent with Fig.~\ref{fig:overall_performance}.

\textbf{Deployment Considerations.} Real-world deployment usually need to consider the following aspects: 1) device-specific profiling to calibrate GPU kernel overheads across driver versions; 2) system-level fault handling (e.g., watchdog, thermal management) to complement failure-aware learning; and 3) inter-cluster coordination for geographically distributed RSU networks, extending beyond the current intra-cluster model ($N=5$). Our implementation supports trace-based calibration for site-specific adaptation, providing a practical foundation for real-world MEC deployment.

\section{Related Work}
\label{sec:related}

\textbf{Latency-Optimal Resource Orchestration.}
Early MEC studies focused on latency minimization via task scheduling and offloading optimization~\cite{kalpana2024deep}. Later work incorporated mobility awareness through Bi-LSTM-based trajectory prediction~\cite{zeng2025task}, collaborative control using advantage actor--critic learning~\cite{liu2023asynchronous, yang2024ha}, and queue-theoretic offloading models regulated by reinforcement learning~\cite{yang2022efficient}. Digital-twin-based designs were further introduced to address uncertainty in resource estimation~\cite{xu2023digital}. Most of these approaches assume homogeneous computing units, which limits their applicability in heterogeneous CPU-GPU infrastructures.

\textbf{Energy-Optimal Resource Management.}
A few studies progressed from device-level power tuning to network-wide coordination through multi-agent reinforcement learning for UAV--MEC networks~\cite{xue2022cost, mondal2024joint}, vehicular offloading with edge caching~\cite{kong2022deep, wu2020toward}, and non-orthogonal multiple access schemes emphasizing transmission efficiency~\cite{chen2024dynamic, ding2019joint}. Physical-layer techniques, such as reconfigurable intelligent surfaces~\cite{zhai2022energy, yang2022intelligent}, further improve energy utilization. However, they decouple communication and computation that comes with suboptimal decisions when transmission power and processing capacity are not jointly optimized.

\textbf{Joint Energy--Latency Optimization.}
Recent work has explored joint energy-latency optimization through game-theoretic modeling~\cite{teng2022game, wu2019energy}, online UAV-assisted control~\cite{dai2023uav, zhang2020energy}, and reinforcement learning under partial observability~\cite{robles2023multi, ale2021delay}. Domain-specific frameworks further consider green-energy-driven offloading~\cite{ma2022greenedge}, renewable-powered vehicular networks~\cite{luo2021minimizing}, and dependency-aware task scheduling~\cite{song2022offloading, xu2023energy}. Nevertheless, unified optimization of continuous power control and discrete task allocation remains challenging, particularly in heterogeneous CPU-GPU environments where task-resource compatibility is critical. To address such limitations, this work proposes a hierarchical learning framework that separates continuous and discrete decision spaces: the upper tier applies TD3 for power control, while the lower tier employs DDQN for task allocation. Coordination mechanisms align both tiers, and GPU-aware compatibility assessment enables efficient resource matching in heterogeneous edge systems.

\section{Conclusion}
\label{sec:conclusion}

This paper investigates resource orchestration for heterogeneous MEC systems characterized by multi-dimensional device heterogeneity, diverse task requirements, and strict timing constraints. Experimental evaluation shows that the proposed hierarchical TD3–DDQN framework attains over 99\% task completion, reduces average latency by 28\%, and maintains balanced energy–performance trade-offs. The framework integrates GPU-aware compatibility assessment, deadline-sensitive queue management, and failure-aware adaptive sampling, enabling stable performance under dynamic and resource-constrained conditions. Future extensions include adapting the compatibility model to new accelerator architectures, exploring predictive task allocation guided by mobility patterns, and validating the hierarchical coordination across broader hardware settings. These directions aim to strengthen the applicability of hierarchical learning frameworks in increasingly heterogeneous edge environments.

\bibliographystyle{IEEEtran}
\bibliography{mybib}

@article{mach2017mobile,
  title={Mobile edge computing: A survey on architecture and computation offloading},
  author={Mach, Pavel and Becvar, Zdenek},
  journal={IEEE communications surveys \& tutorials},
  volume={19},
  number={3},
  pages={1628--1656},
  year={2017},
  publisher={IEEE}
}

@article{zeng2021energy,
  title={Energy-efficient resource management for federated edge learning with CPU-GPU heterogeneous computing},
  author={Zeng, Qunsong and Du, Yuqing and Huang, Kaibin and Leung, Kin K},
  journal={IEEE Transactions on Wireless Communications},
  volume={20},
  number={12},
  pages={7947--7962},
  year={2021},
  publisher={IEEE}
}

@article{lee2022real,
  title={Real-time edge computing on multi-processes and multi-threading architectures for deep learning applications},
  author={Lee, Shih Hsiung},
  journal={Microprocessors and Microsystems},
  volume={92},
  pages={104554},
  year={2022},
  publisher={Elsevier}
}

@inproceedings{kim2020gpu,
  title={GPU-specific task offloading in the mobile edge computing network},
  author={Kim, Namkyu and Lee, Yunseong and Lee, Chunghyun and Nguyen, The Vi and Tuong, Van Dat and Cho, Sungrae},
  booktitle={2020 International Conference on Information and Communication Technology Convergence (ICTC)},
  pages={1874--1876},
  year={2020},
  organization={IEEE}
}

@article{akhlaqi2023task,
  title={Task offloading paradigm in mobile edge computing-current issues, adopted approaches, and future directions},
  author={Akhlaqi, Mohammad Yahya and Hanapi, Zurina Binti Mohd},
  journal={Journal of Network and Computer Applications},
  volume={212},
  pages={103568},
  year={2023},
  publisher={Elsevier}
}

@article{hou2022optimal,
  title={Optimal control of wireless powered edge computing system for balance between computation rate and energy harvested},
  author={Hou, Chen and Zhao, Qianchuan},
  journal={IEEE Transactions on Automation Science and Engineering},
  volume={20},
  number={2},
  pages={1108--1124},
  year={2022},
  publisher={IEEE}
}

@inproceedings{kalpana2024deep,
  title={A deep reinforcement learning-based task offloading framework for edge-cloud computing},
  author={Kalpana, Ponugoti and Almusawi, Muntather and Chanti, Yerrolla and Kumar, V Sunil and Rao, M Varaprasad},
  booktitle={2024 International Conference on Integrated Circuits and Communication Systems (ICICACS)},
  pages={1--5},
  year={2024},
  organization={IEEE}
}

@article{cao2024dependent,
  title={Dependent task offloading in edge computing using GNN and deep reinforcement learning},
  author={Cao, Zequn and Deng, Xiaoheng and Yue, Sheng and Jiang, Ping and Ren, Ju and Gui, Jinsong},
  journal={IEEE Internet of Things Journal},
  volume={11},
  number={12},
  pages={21632--21646},
  year={2024},
  publisher={IEEE}
}

@article{chen2023dynamic,
  title={Dynamic task offloading for digital twin-empowered mobile edge computing via deep reinforcement learning},
  author={Chen, Ying and Gu, Wei and Xu, Jiajie and Zhang, Yongchao and Min, Geyong},
  journal={China Communications},
  volume={20},
  number={11},
  pages={164--175},
  year={2023},
  publisher={IEEE}
}

@article{zeng2025task,
  title={Task offloading delay minimization in vehicular edge computing based on vehicle trajectory prediction},
  author={Zeng, Feng and Zhang, Zheng and Wu, Jinsong},
  journal={Digital Communications and Networks},
  volume={11},
  number={2},
  pages={537--546},
  year={2025},
  publisher={Elsevier}
}

@article{dai2023uav,
  title={UAV-assisted task offloading in vehicular edge computing networks},
  author={Dai, Xingxia and Xiao, Zhu and Jiang, Hongbo and Lui, John CS},
  journal={IEEE Transactions on Mobile Computing},
  volume={23},
  number={4},
  pages={2520--2534},
  year={2023},
  publisher={IEEE}
}

@article{dong2022quantum,
  title={Quantum particle swarm optimization for task offloading in mobile edge computing},
  author={Dong, Shi and Xia, Yuanjun and Kamruzzaman, Joarder},
  journal={IEEE Transactions on Industrial Informatics},
  volume={19},
  number={8},
  pages={9113--9122},
  year={2022},
  publisher={IEEE}
}

@article{mao2016dynamic,
  title={Dynamic computation offloading for mobile-edge computing with energy harvesting devices},
  author={Mao, Yuyi and Zhang, Jun and Letaief, Khaled B},
  journal={IEEE Journal on Selected Areas in Communications},
  volume={34},
  number={12},
  pages={3590--3605},
  year={2016},
  publisher={IEEE}
}

@inproceedings{miettinen2010energy,
  title={Energy efficiency of mobile clients in cloud computing},
  author={Miettinen, Antti P and Nurminen, Jukka K},
  booktitle={2nd USENIX workshop on hot topics in cloud computing (HotCloud 10)},
  year={2010}
}

@article{liu2023asynchronous,
  title={Asynchronous deep reinforcement learning for collaborative task computing and on-demand resource allocation in vehicular edge computing},
  author={Liu, Lei and Feng, Jie and Mu, Xuanyu and Pei, Qingqi and Lan, Dapeng and Xiao, Ming},
  journal={IEEE Transactions on Intelligent Transportation Systems},
  volume={24},
  number={12},
  pages={15513--15526},
  year={2023},
  publisher={IEEE}
}

@article{xu2023digital,
  title={Digital twin-driven collaborative scheduling for heterogeneous task and edge-end resource via multi-agent deep reinforcement learning},
  author={Xu, Chi and Tang, Zixuan and Yu, Haibin and Zeng, Peng and Kong, Linghe},
  journal={IEEE Journal on Selected Areas in Communications},
  volume={41},
  number={10},
  pages={3056--3069},
  year={2023},
  publisher={IEEE}
}

@inproceedings{yang2022efficient,
  title={Efficient resource allocation policy for cloud edge end framework by reinforcement learning},
  author={Yang, Chun and Xu, Hongliu and Fan, Shixiao and Cheng, Xuan and Liu, Minghui and Wang, Xiaomin},
  booktitle={2022 IEEE 8th International Conference on Computer and Communications (ICCC)},
  pages={1363--1367},
  year={2022},
  organization={IEEE}
}

@article{xue2022cost,
  title={Cost optimization of UAV-MEC network calculation offloading: A multi-agent reinforcement learning method},
  author={Xue, Jianbin and Wu, Qingqing and Zhang, Haijun},
  journal={Ad Hoc Networks},
  volume={136},
  pages={102981},
  year={2022},
  publisher={Elsevier}
}

@article{kong2022deep,
  title={Deep reinforcement learning-based energy-efficient edge computing for internet of vehicles},
  author={Kong, Xiangjie and Duan, Gaohui and Hou, Mingliang and Shen, Guojiang and Wang, Hui and Yan, Xiaoran and Collotta, Mario},
  journal={IEEE Transactions on Industrial Informatics},
  volume={18},
  number={9},
  pages={6308--6316},
  year={2022},
  publisher={IEEE}
}

@article{chen2024dynamic,
  title={Dynamic task offloading and resource allocation for NOMA-aided mobile edge computing: An energy efficient design},
  author={Chen, Ying and Xu, Jiajie and Wu, Yuan and Gao, Jie and Zhao, Lian},
  journal={IEEE Transactions on Services Computing},
  volume={17},
  number={4},
  pages={1492--1503},
  year={2024},
  publisher={IEEE}
}

@article{zhai2022energy,
  title={Energy-efficient UAV-mounted RIS assisted mobile edge computing},
  author={Zhai, Zhiyuan and Dai, Xinhong and Duo, Bin and Wang, Xin and Yuan, Xiaojun},
  journal={IEEE wireless communications letters},
  volume={11},
  number={12},
  pages={2507--2511},
  year={2022},
  publisher={IEEE}
}

@article{yang2022intelligent,
  title={Intelligent-reflecting-surface-aided mobile edge computing with binary offloading: Energy minimization for IoT devices},
  author={Yang, Yizhen and Gong, Yi and Wu, Yik-Chung},
  journal={IEEE Internet of Things Journal},
  volume={9},
  number={15},
  pages={12973--12983},
  year={2022},
  publisher={IEEE}
}

@article{robles2023multi,
  title={A multi-layer guided reinforcement learning-based tasks offloading in edge computing},
  author={Robles-Enciso, Alberto and Skarmeta, Antonio F},
  journal={Computer Networks},
  volume={220},
  pages={109476},
  year={2023},
  publisher={Elsevier}
}

@article{teng2022game,
  title={Game theoretical task offloading for profit maximization in mobile edge computing},
  author={Teng, Haojun and Li, Zhetao and Cao, Kun and Long, Saiqin and Guo, Song and Liu, Anfeng},
  journal={IEEE Transactions on Mobile Computing},
  volume={22},
  number={9},
  pages={5313--5329},
  year={2022},
  publisher={IEEE}
}

@article{ma2022greenedge,
  title={GreenEdge: Joint green energy scheduling and dynamic task offloading in multi-tier edge computing systems},
  author={Ma, Huirong and Huang, Peng and Zhou, Zhi and Zhang, Xiaoxi and Chen, Xu},
  journal={IEEE Transactions on Vehicular Technology},
  volume={71},
  number={4},
  pages={4322--4335},
  year={2022},
  publisher={IEEE}
}

@article{luo2021minimizing,
  title={Minimizing the delay and cost of computation offloading for vehicular edge computing},
  author={Luo, Quyuan and Li, Changle and Luan, Tom H and Shi, Weisong},
  journal={IEEE Transactions on Services Computing},
  volume={15},
  number={5},
  pages={2897--2909},
  year={2021},
  publisher={IEEE}
}

@article{song2022offloading,
  title={Offloading dependent tasks in multi-access edge computing: A multi-objective reinforcement learning approach},
  author={Song, Fuhong and Xing, Huanlai and Wang, Xinhan and Luo, Shouxi and Dai, Penglin and Li, Ke},
  journal={Future Generation Computer Systems},
  volume={128},
  pages={333--348},
  year={2022},
  publisher={Elsevier}
}

@article{marzetta2010noncooperative,
  title={Noncooperative cellular wireless with unlimited numbers of base station antennas},
  author={Marzetta, Thomas L},
  journal={IEEE transactions on wireless communications},
  volume={9},
  number={11},
  pages={3590--3600},
  year={2010},
  publisher={IEEE}
}

@article{nasir2025relto,
  title={RELTO: A reliability-oriented DRL approach with context-aware adaptive reward weighting for multi-objective task offloading in MEC},
  author={Nasir, Anam and He, Xiang and Wang, Teng and Shi, Haomei and Wang, Zhongjie},
  journal={Ad Hoc Networks},
  pages={104065},
  year={2025},
  publisher={Elsevier}
}

@article{islam2025delta,
  title={DELTA: Deadline aware energy and latency-optimized task offloading and resource allocation in GPU-enabled, PiM-enabled distributed heterogeneous MEC architecture},
  author={Islam, Akhirul and Ghose, Manojit},
  journal={Journal of Systems Architecture},
  volume={159},
  pages={103335},
  year={2025},
  publisher={Elsevier}
}

@article{wang2024joint,
  title={Joint computation offloading and resource allocation for maritime MEC with energy harvesting},
  author={Wang, Zhen and Lin, Bin and Ye, Qiang and Fang, Yuguang and Han, Xiaoling},
  journal={IEEE Internet of Things Journal},
  volume={11},
  number={11},
  pages={19898--19913},
  year={2024},
  publisher={IEEE}
}

@article{pan2021multi,
  title={A multi-objective clustering evolutionary algorithm for multi-workflow computation offloading in mobile edge computing},
  author={Pan, Lei and Liu, Xiao and Jia, Zhaohong and Xu, Jia and Li, Xuejun},
  journal={IEEE Transactions on Cloud Computing},
  volume={11},
  number={2},
  pages={1334--1351},
  year={2021},
  publisher={IEEE}
}

@article{mondal2024joint,
  title={Joint trajectory, user-association, and power control for green UAV-assisted data collection using deep reinforcement learning},
  author={Mondal, Abhishek and Mishra, Deepak and Prasad, Ganesh and Johansson, H{\aa}kan},
  journal={IEEE Transactions on Intelligent Vehicles},
  year={2024},
  publisher={IEEE}
}

@article{wu2020toward,
  title={Toward energy-aware caching for intelligent connected vehicles},
  author={Wu, Hongjia and Zhang, Jiao and Cai, Zhiping and Liu, Fang and Li, Yangyang and Liu, Anfeng},
  journal={IEEE Internet of Things Journal},
  volume={7},
  number={9},
  pages={8157--8166},
  year={2020},
  publisher={IEEE}
}

@article{ding2019joint,
  title={Joint power and time allocation for NOMA--MEC offloading},
  author={Ding, Zhiguo and Xu, Jie and Dobre, Octavia A and Poor, H Vincent},
  journal={IEEE Transactions on Vehicular Technology},
  volume={68},
  number={6},
  pages={6207--6211},
  year={2019},
  publisher={IEEE}
}

@article{zhang2020energy,
  title={Energy--latency tradeoff for computation offloading in UAV-assisted multiaccess edge computing system},
  author={Zhang, Kaiyuan and Gui, Xiaolin and Ren, Dewang and Li, Defu},
  journal={IEEE Internet of Things Journal},
  volume={8},
  number={8},
  pages={6709--6719},
  year={2020},
  publisher={IEEE}
}

@article{xu2023energy,
  title={Energy consumption and time-delay optimization of dependency-aware tasks offloading for industry 5.0 applications},
  author={Xu, Chen and Lv, Mengzhuo and Zhang, Kun and Cao, Kui and Wang, Gang and Wei, Mingzhu and Peng, Bei},
  journal={IEEE Transactions on Consumer Electronics},
  volume={70},
  number={1},
  pages={1590--1600},
  year={2023},
  publisher={IEEE}
}

@article{ale2021delay,
  title={Delay-aware and energy-efficient computation offloading in mobile-edge computing using deep reinforcement learning},
  author={Ale, Laha and Zhang, Ning and Fang, Xiaojie and Chen, Xianfu and Wu, Shaohua and Li, Longzhuang},
  journal={IEEE Transactions on Cognitive Communications and Networking},
  volume={7},
  number={3},
  pages={881--892},
  year={2021},
  publisher={IEEE}
}

@article{wu2019energy,
  title={Energy-latency aware offloading for hierarchical mobile edge computing},
  author={Wu, Binwei and Zeng, Jie and Ge, Lu and Su, Xin and Tang, Youxi},
  journal={IEEE Access},
  volume={7},
  pages={121982--121997},
  year={2019},
  publisher={IEEE}
}

@article{yang2024ha,
  title={Ha-a2c: Hard attention and advantage actor-critic for addressing latency optimization in edge computing},
  author={Yang, Jing and Lu, Jialin and Zhou, Xu and Li, Shaobo and Xiong, Chuanyue and Hu, Jianjun},
  journal={IEEE Transactions on Green Communications and Networking},
  volume={9},
  number={1},
  pages={207--217},
  year={2024},
  publisher={IEEE}
}
% \end{thebibliography}

% \vskip -3.5\baselineskip
\begin{IEEEbiography}[{\includegraphics[width=1in,height=1.25in,clip,keepaspectratio]{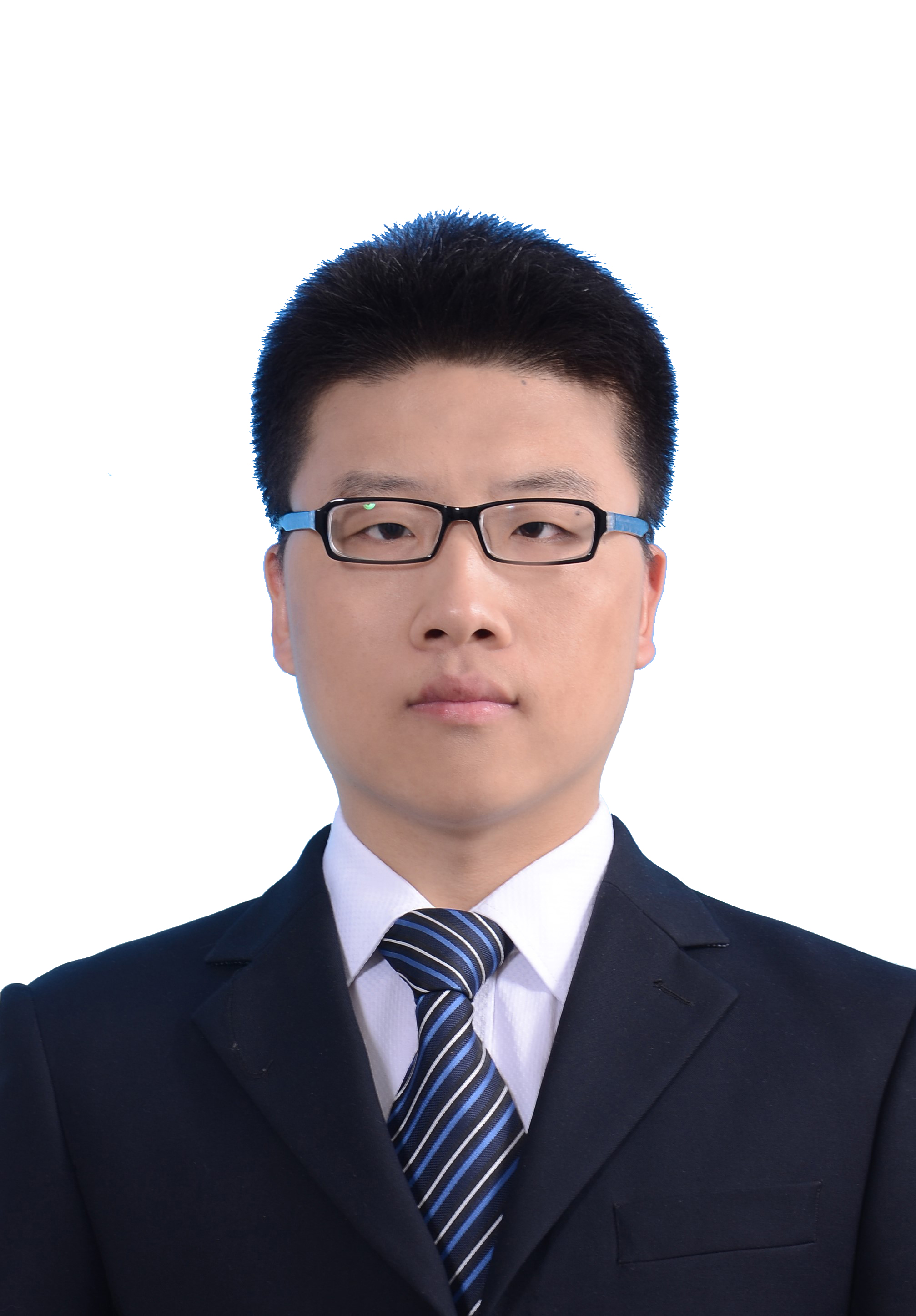}}] {Jianyong Zhu} is currently an assistant professor with the Department of Computing at North China Electric Power University. He received the Ph.D. degree from Beihang University in 2022. His research interests focus on distributed intelligent computing systems, including resource virtualization, learning-enabled resource management, and heterogeneous computing in edge-centric environments.
\end{IEEEbiography}

\vskip -5\baselineskip
\begin{IEEEbiography}[{\includegraphics[width=1in,height=1.25in,clip,keepaspectratio]{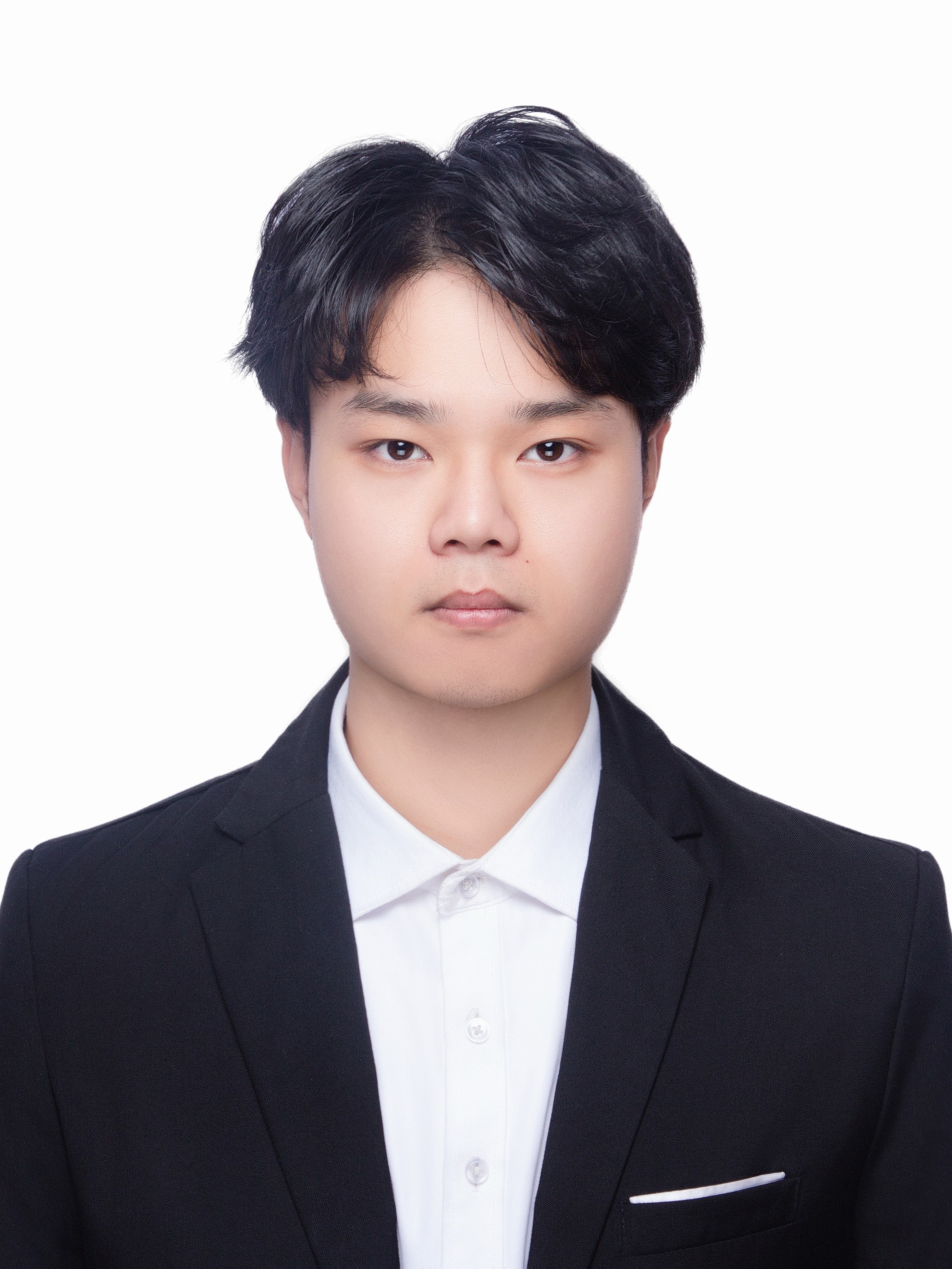}}] {Hao Chen} is currently studying in the Computer Department of North China Electric Power University with a master's degree. His research interests are edge computing, distributed systems and machine learning.
\end{IEEEbiography}

\vskip -5\baselineskip
\begin{IEEEbiography}
[{\includegraphics[width=1in,height=1.25in,clip,keepaspectratio]{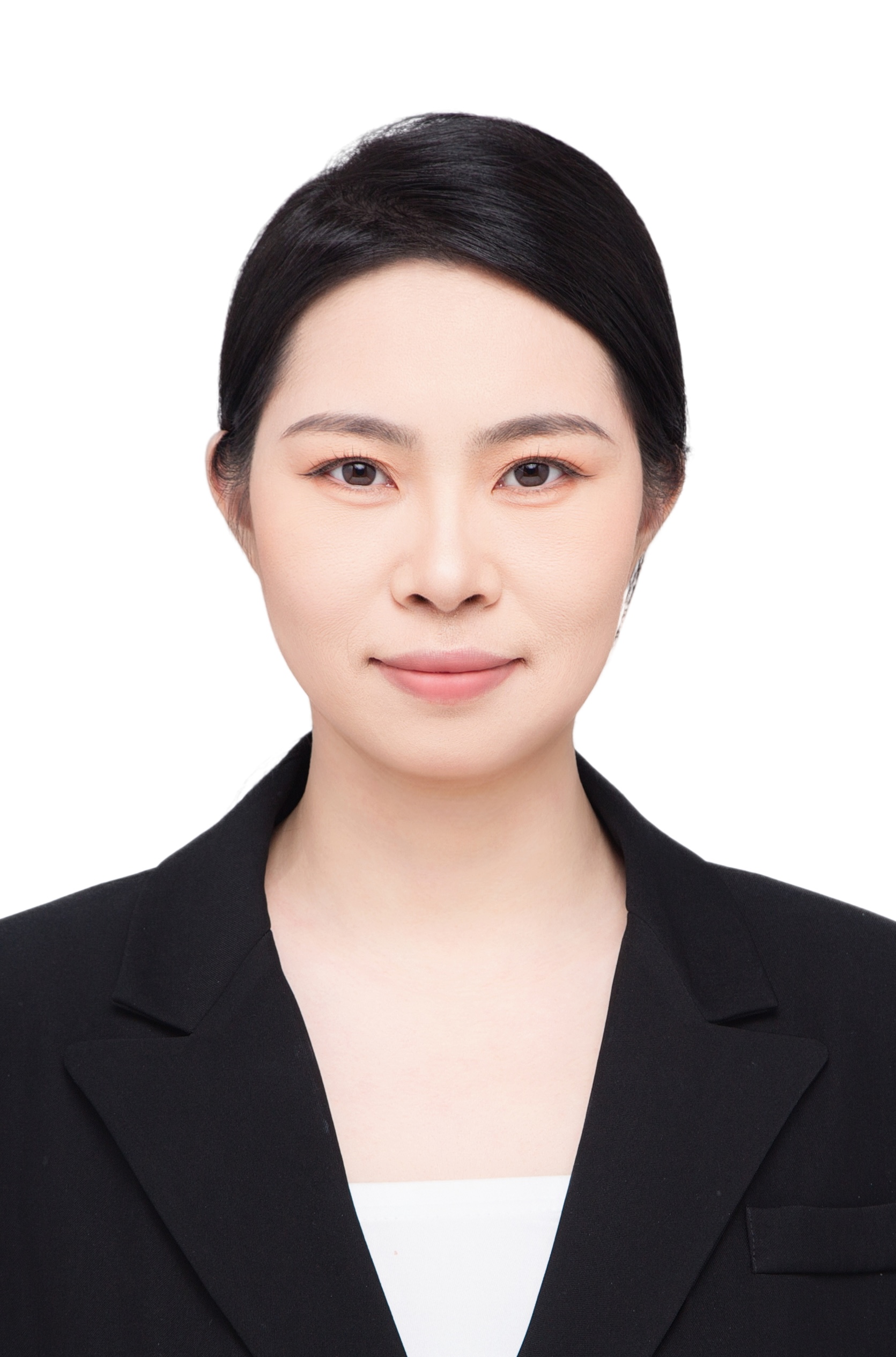}}]{Juan Zhang} (Member, IEEE) is a Lecturer (Assistant Professor) with the James Watt School of Engineering, University of Glasgow, UK. 
%Before this role, she was a Lecturer with the School of Computer Science, Northumbria University at Newcastle, UK, during which she was also a visiting researcher at the Georg-August-University of Göttingen, Germany. 
She received her Ph.D. degree in Computer Science from the University of Exeter, U.K. and subsequently was a postdoctoral researcher with the Chair of High-Performance Computing at Helmut-Schmidt-University/University of the Federal Armed Forces Hamburg, Germany. Her research interests include mobile edge computing, decision-making strategies, IoT, and future mobile networks.
\end{IEEEbiography}

\vskip -5\baselineskip
\begin{IEEEbiography}[{\includegraphics[width=1.1in,height=1.25in,clip,keepaspectratio]{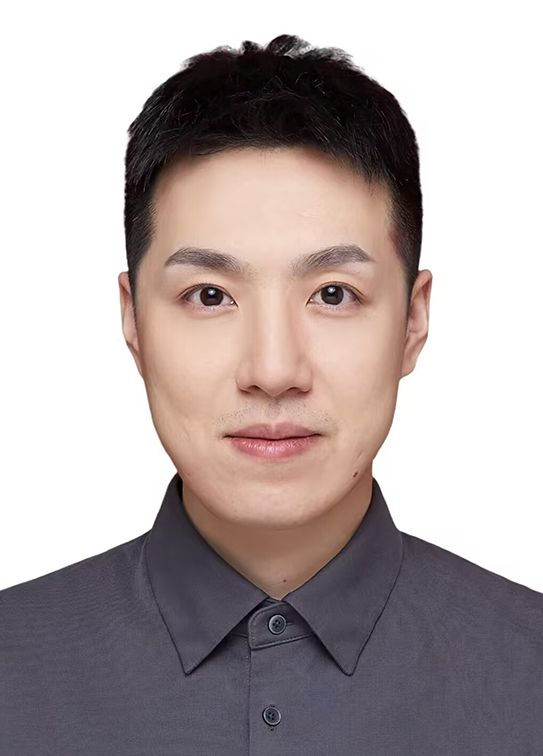}}]{Fangda Guo} is an assistant professor with the Institute of Computing Technology, Chinese Academy of Sciences. He received the Ph.D. degree from Northeastern University, China, in 2021. He was a joint Ph.D. candidate with the University of Edinburgh, UK.  His research interests include social computing, data mining, graph management, query optimization and AI safety. \end{IEEEbiography}

\vskip -5\baselineskip
\begin{IEEEbiography}
[{\includegraphics[width=1in,height=1.25in,clip,keepaspectratio]{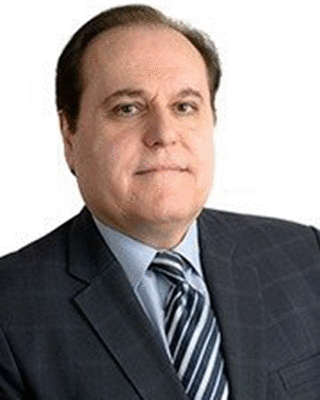}}]{Albert Y. Zomaya} (Fellow, IEEE) is the Peter Nicol Russell chair professor of computer science with the School of Computer Science, Sydney University, and serves as the director of the Centre for Distributed and High-Performance Computing. He has published more than 800 scientific papers and articles and is the author, co-author, or editor of more than 30 books. He is the past editor-in-chief of the \textit{IEEE Trans on Computers},the \textit{IEEE Trans on Sustainable Computing}, and the \textit{ACM Computing Surveys}. His research interests include parallel and distributed computing, networking, and complex systems.
\end{IEEEbiography}

\vskip -5\baselineskip
\begin{IEEEbiography}
[{\includegraphics[width=1in,height=1.25in,clip,keepaspectratio]{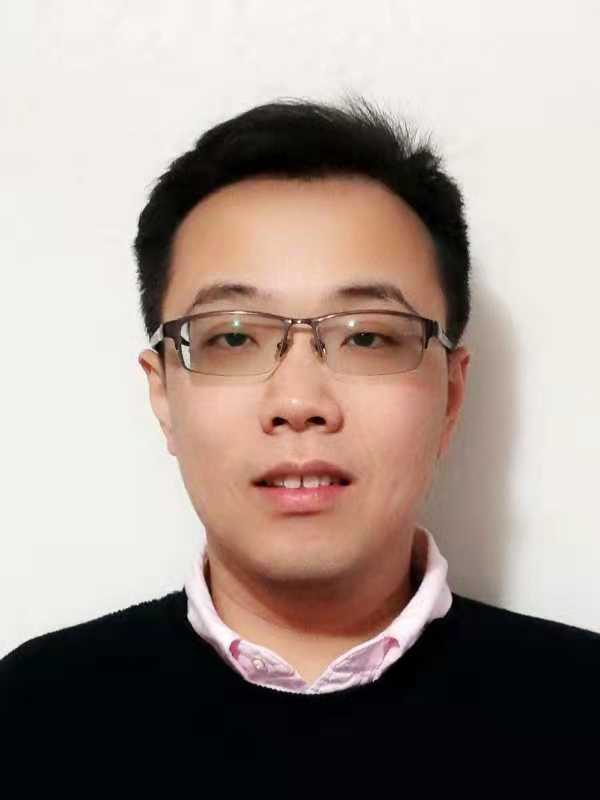}}]{Renyu Yang} (Member, IEEE) is an associate professor with the School of Software, Beihang University, China. Prior to this, he was with the University of Leeds UK, Alibaba Group China and Edgetic Ltd. UK, building large-scale computing/AI infrastructures. He is a recipient of Alan Turing Post-Doctoral Enrichment Award, 2022. His research interests include parallel and distributed computing, large-scale AI systems and software dependability. 
\end{IEEEbiography}
\end{document}